\documentclass[twocolumn,twocolappendix]{aastex701}

\shorttitle{{\sc Effortless} II. Implementation}
\shortauthors{K. Cao}
\usepackage{CJKutf8}

\usepackage{graphicx}	
\usepackage{amsmath}	
\usepackage{amssymb}	
\usepackage{gensymb}
\usepackage{multirow}

\graphicspath{{figures/}}
\usepackage[OT2,T1]{fontenc}
\DeclareSymbolFont{cyrletters}{OT2}{wncyr}{m}{n}
\DeclareMathSymbol{\Sha}{\mathalpha}{cyrletters}{"58}

\begin{document}

\title{Efficient Optimal Image Reconstruction for the Nancy Grace Roman Space Telescope and Beyond: II. Implementation of {\sc Effortless}}

\author[orcid=0000-0002-1699-6944]{Kaili Cao (\begin{CJK*}{UTF8}{gbsn}曹开力\end{CJK*})}
\affiliation{Center for Cosmology and AstroParticle Physics (CCAPP), The Ohio State University, 191 West Woodruff Ave, Columbus, OH 43210, USA}
\affiliation{Department of Physics, The Ohio State University, 191 West Woodruff Ave, Columbus, OH 43210, USA}
\email[show]{cao.1191@osu.edu}

\collaboration{all}{Roman HLIS Cosmology PIT}

\begin{abstract}

Weak gravitational lensing is a promising but technically demanding cosmological probe. For space missions like the forthcoming Nancy Grace Roman Space Telescope, a major challenge is that native images are undersampled and need to be reconstructed to enable accurate measurements. {\sc Effortless} (EFFicient Optimal image ReconsTruction using LESS memory; previously known as Fast {\sc Imcom}) is a new algorithm designed for that purpose. My companion paper has exhibited promising first results to demonstrate that {\sc Effortless} can make point spread functions (PSFs) uniform and regular across reconstructed images more efficiently than its predecessor {\sc Imcom} and has the potential to outperform {\sc Imcom} in terms of control over systematic errors. In this paper, I present the mathematical formalism, software implementation, and practical issues in detail. Foremost, while the Nyquist--Shannon sampling theorem remains true, the conditions of the theorem are subtly (and importantly) different from the problem in survey data processing, and finite sampling effects can be substantially reduced via a simple post-measurement calibration. Imperfections caused by numerical artifacts, finiteness of input pixel windows, and unavailability of some input pixels are understood and under control. The {\sc Effortless} application programming interface is general and can support use cases beyond weak lensing cosmology and beyond the Roman Space Telescope.

\end{abstract}

\keywords{\uat{Astronomy image processing}{2306} --- \uat{Weak gravitational lensing}{1797}}	

\section{Introduction} \label{sec:intro}

Massive objects gravitate. In addition to attracting each other, they also distort light paths in their vicinities. Such behavioral pattern, termed gravitational lensing, allows us to probe the large-scale structure of the Universe without being concerned by the incompleteness of our knowledge about how galaxies trace the matter distribution. The scenario where the shapes of distant galaxies are only distorted but not replicated is referred to as weak gravitational lensing \citep[see][for some recent reviews]{2013PhR...530...87W, 2015RPPh...78h6901K, 2018ARA&A..56..393M}. While weak lensing is a promising cosmological probe, it is also technically challenging, since intrinsic shapes of background galaxies differ at the $\sim 30\%$ level, while the changes caused by foreground objects are only at the $< 1\%$ level. Therefore, measurement of cosmic shear imposes stringent requirements on image processing (see, e.g., the Roman Space Telescope Science Requirements Document\footnote{\url{https://roman.gsfc.nasa.gov/science/docs/RST-SYS-REQ-0020D_DOORs_Export.pdf}}).

Despite the difficulty, the field of weak lensing cosmology has advanced significantly over the past few decades. The ``Stage III'' surveys \citep[as defined in][]{2006astro.ph..9591A} --- the Kilo Degree Survey \citep[KiDS;][]{2022A&A...664A.170V, 2023A&A...669A..69B, 2023A&A...675A.189D, 2023A&A...679A.133L, 2025A&A...703A.158W}, the Dark Energy Survey \citep[DES;][]{2022PhRvD.105b3514A, 2022PhRvD.105b3515S, 2026arXiv260114559D, 2026arXiv260210065D, 2026PhRvD.113j3503G}, and the Hyper Suprime Cam \citep[HSC;][]{2019PASJ...71...43H, 2020PASJ...72...16H, 2023PhRvD.108l3517M, 2023PhRvD.108l3518L, 2023PhRvD.108l3519D, 2023PhRvD.108l3520M, 2023PhRvD.108l3521S} --- have tightened constraints on values of cosmological parameters to the percent level. The upcoming or ongoing ``Stage IV'' surveys --- NASA's Nancy Grace Roman Space Telescope \citep[hereafter Roman;][]{2019arXiv190205569A, 2025arXiv250510574O}, the Legacy Survey of Space and Time (LSST) at the NSF-DOE Vera C. Rubin Observatory \citep{2012arXiv1211.0310L, 2019ApJ...873..111I}, and the Euclid space telescope \citep{2011arXiv1110.3193L, 2022A&A...662A.112E, 2025A&A...697A...1E} --- promise to push the cosmological uncertainties down to the sub-percent level.

For space missions like Roman, the vantage point of outer space is a blessing and a curse in terms of point spread functions (PSFs). On the one hand, the PSFs are stable and can be modeled by tools like STPSF for Roman\footnote{\url{https://roman-docs.stsci.edu/simulation-tools-handbook-home/stpsf-for-roman}}; in demanding cases like weak lensing, the models can be iteratively corrected using {\sc PIFF} \citep{2021MNRAS.501.1282J}. On the other hand, the PSFs are narrow and thus the native images are ``undersampled.'' An astronomical image is said to be oversampled, Nyquist sampled, or undersampled if its pixel size is less than, equal to, or greater than half its PSF size; the PSF size is usually characterized by the full width at half maximum (FWHM). State-of-the-art shear measurement algorithms like {\sc Metacalibration} \citep{2017arXiv170202600H, 2017ApJ...841...24S}, {\sc Metadetection} \citep{2020ApJ...902..138S}, and {\sc AnaCal} \citep[also known as {\sc FPFS};][]{2023MNRAS.521.4904L, 2024MNRAS.52710388L} operate on oversampled images. For ground-based facilities, images are typically oversampled because of seeing conditions of the Earth's atmosphere; for space-based instruments like Roman, the PSFs are diffraction-limited and have widths $\sim \lambda/D$, where $\lambda$ denotes the wavelength of observation and $D$, the entrance pupil diameter.

Out of survey efficiency considerations, the pixel size of the Roman Wide Field Instrument \citep[WFI;][]{2020JATIS...6d6001M} is $0.11 \,{\rm arcsec}$, which is comparable to PSF sizes in near infrared bands. Since the spatial resolution needs to be about twice as high to reach Nyquist sampling, native Roman images are undersampled and should be reconstructed before being passed to shear calibration pipelines. In their Appendix~C, \citet{2024MNRAS.528.2533H} argued that several undersampled images need to be combined to resolve aliasing during the reconstruction process. Such belief underpins the design of the Roman High Latitude Imaging Survey (HLIS), imaging component of High-Latitude Wide-Area Survey \citep[HLWAS; see Appendix~C of][]{2025arXiv250510574O}: For most of the survey regions, each pointing will be covered by $3$ dithers $\times 2$ passes $= 6$ exposures (the actual coverage can be slightly smaller due to detector gaps and bad pixels) in each band. While such design is advisable in other regards as well, the image reconstruction problem in survey data processing can be subtly (and importantly) different from the conditions of the Nyquist--Shannon sampling theorem. As I demonstrate in this series of papers, given full information about native PSFs, it is possible to reconstruct oversampled images from individual native images and enable accurate measurements.

Image reconstruction is usually formulated as linear transformations \citep[see][for the motivation]{2023OJAp....6E...5M}. A variety of software tools have been developed for this purpose \citep[e.g.,][]{1999PASP..111..227L, 2017ApJ...836..187Z, 2017ApJ...836..188Z}, among which {\sc Drizzle} \citep{2002PASP..114..144F, 2012drzp.book.....G} is widely used, largely because of its ability to handle arbitrary dithers. However, {\sc Drizzle} outputs have neither uniform and regular PSFs nor readily comprehensible noise patterns, both of which are necessary for precision weak lensing cosmology. By contrast, {\sc Imcom} \citep[IMage COMbination;][]{2011ApJ...741...46R}, which can also handle arbitrary dithers, provides control over both output PSFs and noise amplification. Therefore, it is currently adopted by the Roman Project Infrastructure Team (PIT) ``Maximizing Cosmological Science with the Roman High Latitude Imaging Survey'' (PI: O. Dor\'e)\footnote{\url{https://roman-hlis-cosmology.caltech.edu/}} for image reconstruction. My colleagues and I have been developing {\sc Imcom} \citep{2024MNRAS.528.2533H, 2024MNRAS.528.6680Y, 2025ApJS..277...55C, 2026ApJ...998..304C} and validating it using image simulations \citep{2023MNRAS.522.2801T, 2025MNRAS.544.3799O}. It is our expectation that minimal systematics in shear measurement will translate into higher cosmological yields \citep[e.g.,][]{2026arXiv260100438C}.

As a successor to {\sc Imcom}, {\sc Effortless} \citep[previously known as Fast {\sc Imcom};][]{2026AJ....171..140C} has the same optimization goals. Leveraging the knowledge about patterns manifested by {\sc Imcom} solutions, {\sc Effortless} can produce images of higher quality with $\lesssim 2\%$ of computational resources \citep{2026arXiv260706674C}. Furthermore, since {\sc Effortless} has the capability of constructing oversampled images using individual native images, it can greatly facilitate comparisons between single-epoch images and thus support, e.g., time-domain sciences with Roman within \citep{2019ApJS..241....3P, 2023ApJS..269....5W, 2025ApJ...987..181W} and beyond \citep{2025ApJ...988...65R, 2025ApJ...993..116K} the Milky Way. To be commensurate with its role as a general-purpose astronomical image processing tool, {\sc Effortless} needs to meet higher software design requirements than {\sc PyImcom} \citep[current implementation of {\sc Imcom};][]{2025ApJS..277...55C}. As presented in this paper, the {\sc Effortless} software has a more convenient object-oriented framework, a more general and flexible data model, and a more user-friendly I/O interface. With these features, {\sc Effortless} can simultaneously benefit studies of static and dynamic aspects of the Universe.

This paper is structured as follows. In Section~\ref{sec:math}, I review the mathematical formalism of the {\sc Effortless} algorithm. In Section~\ref{sec:software}, I present the implementation of the {\sc Effortless} software to provide the necessary context for the rest of the paper. In Section~\ref{sec:residual}, I examine three major sources of PSF residuals and tune corresponding hyperparameters for simulations in \citet{2026arXiv260706674C} and this work. Then in Section~\ref{sec:mask}, I sequester and mitigate imperfections due to masked input pixels. In Section~\ref{sec:calibr}, I motivate and conduct the post-measurement calibration. Finally in Section~\ref{sec:disc}, I summarize the main findings of this paper and discuss their implications.

\section{Mathematical Formalism} \label{sec:math}

The mathematical formalism of the {\sc Effortless} algorithm was introduced in \citet[][sometimes referred to as the {\sc Effortless} formalism paper]{2026AJ....171..140C}, along with in-depth discussions of underlying concepts. This section is a review from a practical perspective; it focuses on ``how'' {\sc Effortless} reconstructs images and only explains ``why'' when necessary. In addition to what was in the formalism paper, it provides a roadmap to developments in Sections~\ref{sec:residual} to \ref{sec:calibr} (Section~\ref{ss:math_gen}) and addresses the practical issue of geometric distortions (Section~\ref{ss:distort}). Throughout this paper, I only discuss the reconstruction of individual images; see the formalism paper for how {\sc Effortless} combines multiple images.

\subsection{General Formalism} \label{ss:math_gen}

All linear image reconstruction schemes (including but not limited to {\sc Drizzle}, {\sc Imcom}, and {\sc Effortless}) are linear transformations from input signals $I_{\boldsymbol i}$ to output signals $H_{\boldsymbol \alpha}$:
\begin{equation}
    H_{\boldsymbol \alpha} = \sum_{{\boldsymbol i} \in {\mathbb Z}^2} T_{{\boldsymbol \alpha} {\boldsymbol i}} I_{\boldsymbol i},
    \label{eq:transform}
\end{equation}
where ${\boldsymbol i}$ and ${\boldsymbol \alpha}$ are input and output pixel indices, respectively; both are $2$-tuples of integers since astronomical images are 2D. Different algorithms differ because of their different ways of assigning the reconstruction weights $T_{{\boldsymbol \alpha} {\boldsymbol i}}$: {\sc Drizzle} uses geometric overlaps between (shrunk) input pixels and output pixels, {\sc Imcom} builds and solves linear systems to minimize PSF residuals (see below), and {\sc Effortless} emulates (without using machine learning) what {\sc Imcom} does. What is in common is that they only assign non-zero weights to input pixels that are adjacent and available. The locality is mandated by the finiteness of computational resources and has been found to meet practical needs; making lost input pixels available via interpolation (i.e., remedying input images) is beyond the scope of this paper, which focuses on the reconstruction process. See Section~\ref{ss:accept} for what ``adjacent'' means in this context and Section~\ref{sec:mask} for how {\sc Effortless} handles unavailability of some input pixels in the vicinity of each output pixel.

The next question is how to interpret the signals: What does it mean when we say input pixel ${\boldsymbol i}$ has signal $I_{\boldsymbol i}$? From an instrumental point of view, this simply means that input pixel ${\boldsymbol i}$ gives us some digital number, which is then converted to an analog signal and calibrated to yield $I_{\boldsymbol i}$. However, if we have an understanding of the point spread function (PSF), we can ask a follow-up question: Where does the signal $I_{\boldsymbol i}$ come from? In \citet{2026AJ....171..140C}, I introduced two related but different concepts (with $ {\boldsymbol s}$ denoting the relative position): A ``forward'' PSF $G ({\boldsymbol s})$ is the probability distribution of the landing location of a photon given its source direction, and a ``backward'' PSF $G' ({\boldsymbol s})$ is the probability distribution of the source direction of a photon given its landing location. While we can measure forward PSFs from images of (bright but unsaturated) stars, the physical meaning of input signals is described by backward PSFs: When we attribute the signal in input pixel ${\boldsymbol i}$ to points on the celestial sphere (usually in its vicinity), $I_{\boldsymbol i}$ determines the total flux to be attributed, and $G' ({\boldsymbol s})$ encodes how that flux should be distributed. Without pixelation, $G ({\boldsymbol s})$ and $G' ({\boldsymbol s})$ are just the inversion of each other, as can be proven using Bayes' theorem.\footnote{For instance, if a 1D forward PSF $G (s)$ has a maximum at $s = 0$ but a negative mean, photons from a point source are more likely to be detected to its left than its right. Then if we detect a photon somewhere, it is more likely to have come from its right than its left, i.e., the corresponding backward PSF $G' (s)$ has a positive mean.} The formalism paper was missing such inversion (it did not matter as the input PSFs therein were symmetric), but readers are still referred to that paper for how pixelation is incorporated. Combining these two aspects, a backward PSF $G' ({\boldsymbol s})$ is the convolution of the inverted version of the corresponding forward PSF $G (-{\boldsymbol s})$ and the pixelation function $\Pi ({\boldsymbol s})$, which is often assumed to be a 2D top-hat.

With the physical meaning of input signals $I_{\boldsymbol i}$ elucidated, that of output signals $H_{\boldsymbol \alpha}$ also becomes clear. By assigning weights to input pixels, Equation~(\ref{eq:transform}) also constructs a backward output PSF $\Psi'_{\boldsymbol \alpha}$\footnote{In \citet{2026AJ....171..140C}, I omitted the prime $'$ for ``backward'' output PSFs, as I did not talk much about ``forward'' output PSFs there.} from properly shifted backward input PSFs $G'_{\boldsymbol i}$:
\begin{equation}
    \Psi'_{\boldsymbol \alpha} ({\boldsymbol s}) = \sum_{{\boldsymbol i} \in {\mathbb Z}^2} T_{{\boldsymbol \alpha} {\boldsymbol i}} G'_{\boldsymbol i} ({\boldsymbol R}_{\boldsymbol \alpha} - {\boldsymbol r}_{\boldsymbol i} + {\boldsymbol s}),
    \label{eq:recons_pixel}
\end{equation}
where ${\boldsymbol R}_{\boldsymbol \alpha}$ and ${\boldsymbol r}_{\boldsymbol i}$ are the positions of output pixel ${\boldsymbol \alpha}$ and input pixels ${\boldsymbol i}$, respectively. (If a position is at ${\boldsymbol s}$ relative to ${\boldsymbol R}_{\boldsymbol \alpha}$, it is at ${\boldsymbol R}_{\boldsymbol \alpha} - {\boldsymbol r}_{\boldsymbol i} + {\boldsymbol s}$ relative to ${\boldsymbol r}_{\boldsymbol i}$.) Both {\sc Imcom} and {\sc Effortless} aim at minimizing the discrepancy between Equation~(\ref{eq:recons_pixel}) and some user-specified target output PSF $\Gamma'$; mathematically, the ``PSF leakage'' metric is defined as:
\begin{equation}
    \frac{U_{\boldsymbol \alpha}}{C} \equiv \frac{\Vert \Psi'_{\boldsymbol \alpha} - \Gamma' \Vert^2}{\Vert \Gamma' \Vert^2},
    \label{eq:leakage}
\end{equation}
where $\Vert \cdot \Vert$ is the $L^2$ norm; note that $C \equiv \Vert \Gamma' \Vert^2$ is a constant for a given target output PSF $\Gamma'$, which is usually chosen as a symmetric bivariate Gaussian \citep{2026ApJ...998..304C}:
\begin{equation}
    \Gamma' ({\boldsymbol s}) = \frac{\exp[-{\boldsymbol s}^2/(2\sigma^2)]}{2\pi\sigma^2},
    \label{eq:gauss}
\end{equation}
where $\sigma$ is referred to as the ``target output PSF width'' throughout this paper; it needs to ensure that $\Gamma'$ is at least slightly wider than input PSFs.

{\sc Effortless} makes the reasonable approximation that, in the vicinity of a given position, $G'$ is the same for all input pixels, i.e., $G'_{\boldsymbol i}$ can be replaced by $G'$. Under this approximation, {\sc Effortless} computes reconstruction weights for input pixels by sampling the weight field $T$:
\begin{equation}
    T_{{\boldsymbol \alpha} {\boldsymbol i}} = T ({\boldsymbol r}_{\boldsymbol i} - {\boldsymbol R}_{\boldsymbol \alpha}), \quad {\tilde T} = {\tilde \Gamma}' / {\tilde G}',
    \label{eq:T_solution}
\end{equation}
where ${\tilde \cdot}$ denotes Fourier transform. As one would imagine, there are two issues involved here. First, both ${\tilde \Gamma}'$ and ${\tilde G}'$ vanish (at least in computers) at large wavenumbers, hence the division in Fourier space leads to numerical artifacts; mitigation of such artifacts is discussed in Section~\ref{ss:limit}. Second, because the sampling is finite (we can only assign weights to where the input pixels are located), the resulting output PSFs $\Psi'_{\boldsymbol \alpha}$ are not exactly the same as the target output PSF $\Gamma'$; the corresponding PSF residuals are characterized in Section~\ref{ss:samp}, and calibration of their impact on measurements is motivated and conducted in Section~\ref{sec:calibr}.

\subsection{Geometric Distortions} \label{ss:distort}

Another important topic is the handling of geometric distortions. Astronomical images are flat, while the celestial sphere is not, and the telescope optics induce complicated distortion patterns; nevertheless, in small areas of the sky, the deviations are small, and the coordinate transformations can usually be sufficiently described by world coordinate systems (WCSes) written as fourth- or fifth-order polynomials. The problem is that input and output images have different WCSes, and a choice needs to be made between computing Equation~(\ref{eq:T_solution}) in the input pixel plane or the output pixel plane. For PSF-related calculations, {\sc PyImcom} projects both input pixels and input PSFs onto the output pixel plane; however, {\sc Effortless} treats input images as regular arrays of pixels, therefore it is natural to work in input pixel planes. To achieve this goal, {\sc Effortless} captures linear terms in the coordinate transformations between input and output pixel planes via the Jacobian matrices, referred to as distortion matrices in this context:
\begin{align}
    \nonumber {\bf D} &\equiv \begin{pmatrix} \partial x_{\rm out} / \partial x_{\rm in} & \partial x_{\rm out} / \partial y_{\rm in} \\
    \partial y_{\rm out} / \partial x_{\rm in} & \partial y_{\rm out} / \partial y_{\rm in} \end{pmatrix} \\
    &= \begin{pmatrix} \partial \alpha / \partial x_{\rm out} & \partial \alpha / \partial y_{\rm out} \\
    \partial \delta / \partial x_{\rm out} & \partial \delta / \partial y_{\rm out} \end{pmatrix}^{-1}
    \begin{pmatrix} \partial \alpha / \partial x_{\rm in} & \partial \alpha / \partial y_{\rm in} \\
    \partial \delta / \partial x_{\rm in} & \partial \delta / \partial y_{\rm in} \end{pmatrix},
    \label{eq:distort}
\end{align}
where $(x_{\rm out}, y_{\rm out})^{\rm T}$ and $(x_{\rm in}, y_{\rm in})^{\rm T}$ are representations of the same position or displacement vector in output and input pixel planes, respectively; $\alpha$ and $\delta$ are right ascension and declination in the equatorial coordinate system. Note that $(x_{\rm out}, y_{\rm out})^{\rm T}$ and $(x_{\rm in}, y_{\rm in})^{\rm T}$ are supposed to have the same units; in practice, this means scaling factors may need to be included when computing Equation~(\ref{eq:distort}).

For example, the relative position vector ${\boldsymbol s}$ transforms as ${\boldsymbol s}_{\rm out} = {\bf D} {\boldsymbol s}_{\rm in}$ and ${\boldsymbol s}_{\rm in} = {\bf D}^{-1} {\boldsymbol s}_{\rm out}$. Since we want a uniform PSF in the output pixel plane, the ${\boldsymbol s}$ in Equation~(\ref{eq:gauss}) should be ${\boldsymbol s}_{\rm out}$; in the input pixel plane, it needs to be rewritten as:
\begin{equation}
    \Gamma'_{\rm in} ({\boldsymbol s}_{\rm in}) = \frac{\exp[-({\bf D} {\boldsymbol s}_{\rm in})^2/(2\sigma^2)]}{2\pi\sigma^2},
    \label{eq:gauss_in}
\end{equation}
where $\Gamma'_{\rm in}$ (with the subscript ``in'') means the target output PSF $\Gamma'$ represented in the input pixel plane. Note that as it is written, Equation~(\ref{eq:gauss_in}) is not normalized to unity (in terms of the total integral); this is a feature, not a bug: If we want $\Gamma'$ to be normalized to unity in the output pixel plane, then its representation in the input pixel plane should be normalized to $|{\bf D}|^{-1}$, so that the flux is conserved. When {\sc Effortless} computes Equation~(\ref{eq:T_solution}), the input PSF $G'$ is usually provided in the input pixel plane; it uses Equation~(\ref{eq:gauss_in}) instead of Equation~(\ref{eq:gauss}), and samples the resulting weight field $T$ in the input pixel plane. Equation~(\ref{eq:transform}) is discrete, and $T_{{\boldsymbol \alpha} {\boldsymbol i}}$ values obtained in this way should work equally well.

\begin{figure*}[t!]
    \centering
    \includegraphics[width=\textwidth]{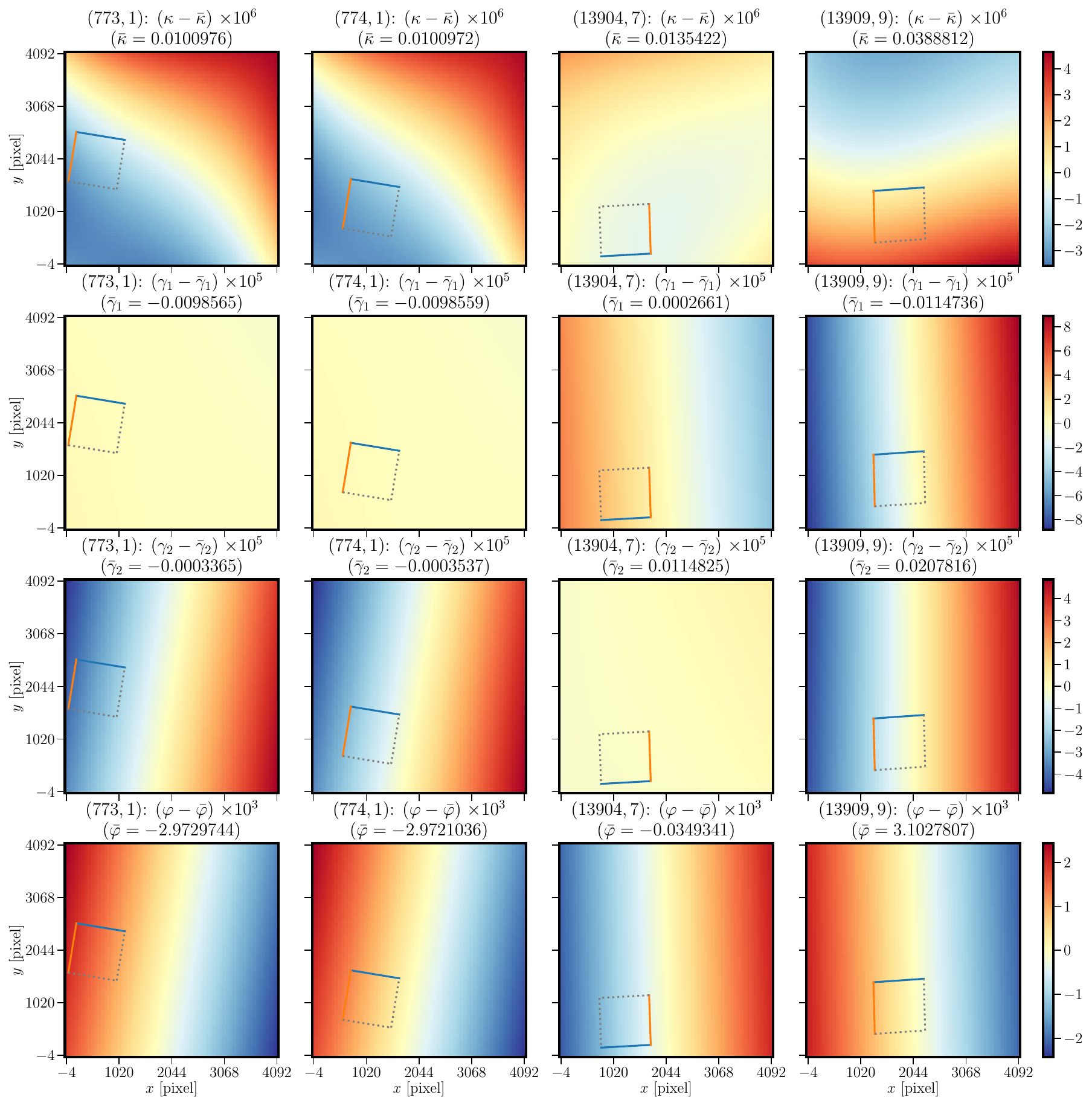}
    \caption{Decomposition of distortion matrices. From top to bottom, the four rows present the four components of distortion matrices, the convergence $\kappa$, two components of the shear $\gamma_1$ and $\gamma_2$, and the roll angle $\varphi$, respectively. The four columns correspond to a representative set of four input images; while these are simulated images in the H158 band, the distortion matrices are band-independent. Each panel visualizes the spatial variation of a distortion component for an image by plotting the discrepancies between values at different positions (in input pixel coordinates; the domain is $[-4, 4092]^2$ to include reference pixels) and the average value in that panel. In each row, the discrepancies are multiplied by some power of $10$ to avoid clutter; average values (denoted by bars) are annotated on top of the panels. The four chosen images all overlap with block $(9, 20)$ in this work, whose $x$- and $y$-axes are shown as orange and blue solid line segments in each panel, while its other two edges are shown as gray dotted line segments.
    \label{fig:distort_mat}}
\end{figure*}

To develop an understanding of real-world distortion matrices, we can decompose them following Equation~(C4) of \citet{2024MNRAS.528.2533H}:
\begin{equation}
    {\bf D}_{\kappa, {\boldsymbol \gamma}, \varphi} = \begin{pmatrix} 1 - \kappa - \gamma_1 & - \gamma_2 \\ - \gamma_2 & 1 - \kappa + \gamma_1 \end{pmatrix}
    \begin{pmatrix} \cos \varphi & - \sin \varphi \\ \sin \varphi & \cos \varphi \end{pmatrix},
    \label{eq:decomp}
\end{equation}
where $\kappa$ is the convergence, ${\boldsymbol \gamma}$ is the $2$-component shear, and $\varphi \in (-\pi, \pi]$ is the roll angle. All nonsingular $2 \times 2$ matrices can be decomposed in this way. Since the distortions are small, it is expected that $\kappa, |{\boldsymbol \gamma}| \ll 1$. Figure~\ref{fig:distort_mat} presents the decomposition of distortion matrices from the combination of four OpenUniverse2024 \citep{2025MNRAS.544.3799O} simulated images and one of the blocks (a block is an output image in the {\sc Imcom} terminology) processed in \citet{2026ApJ...998..304C} and this series of papers. Note that each input image is numbered by the observation ID (e.g., $773$) and the sensor chip assembly (SCA) ID (e.g., $1$). This set of four input images is deliberately chosen to be representative: Two of them are from the first pass of the survey footprint, while the other two are from the second pass; furthermore, SCA $1$ is near the center of the Roman WFI, and SCAs $7$ and $9$ are at the corners of WFI. For a diagram of the WFI layout, see the Roman Documentation (RDox).\footnote{\url{https://roman-docs.stsci.edu/roman-instruments/the-wide-field-instrument/description-of-the-wfi}}

The size of a Roman SCA is $(7.5 \,{\rm arcmin})^2$ on the celestial sphere, which is relatively large by the standards of high-resolution space telescopes in optical and near infrared. Nevertheless, the spatial variations of the four components of Equation~(\ref{eq:decomp}) are still very small: at the $10^{-6}$ level for $\kappa$, at the $10^{-5}$ level or below for both components of ${\boldsymbol \gamma}$, and at the $10^{-3}$ level (in units of radians) for $\varphi$. In this series of papers, {\sc Effortless} samples the distortion matrices along with input PSFs every $(2.5 \,{\rm arcsec})^2$; since the spatial variations of the former are small, this sampling rate is mainly motivated by the spatial variations of the latter. For a given pointing, while the roll angles also depend on the seasons of the observations, the convergence and the shear are largely intrinsic properties for the WFI configuration, as evidenced by the similarity between the first two columns of Figure~\ref{fig:distort_mat}.
Looking at the $\bar \kappa$ and $\bar {\boldsymbol \gamma}$ values annotated on top of the panels, we can see that these numbers are indeed $\ll 1$, but they still need to be taken into account for precise measurements. Also shown in the figure is the chosen block, whose size is $(1.75 \,{\rm arcmin})^2$; from its different orientations in different columns, we can also see the roll angles of different observations. Note that observations $13904$ and $13909$ are two dithers of the same pass, but the roll angles in the last two columns differ by $\sim \pi$, because detector coordinate systems of SCAs $3n$ ($n \in {\mathbb Z} \cap [1, 6]$) are ``rotated'' $180 \degree$ compared to those of SCAs $3n-2$ and $3n-1$; also see the RDox.\footnote{\url{https://roman-docs.stsci.edu/simulation-tools/additional-simulation-tools/pysiaf-for-roman}}

\section{Software Implementation} \label{sec:software}

This section is a narrative that helps develop intuition about how {\sc Effortless} works; it is not a comprehensive documentation of the package. I describe the software structure and the hyperparameters in Sections~\ref{ss:struct} and \ref{ss:params}, respectively.

\subsection{Software Structure} \label{ss:struct}

Object-oriented programming (OOP) was advertised as a feature of {\sc PyImcom} \citep{2025ApJS..277...55C}. Indeed, it makes extensive use of classes to manage data in an intuitive manner. But apart from that, {\sc PyImcom} hardly provides a general application programming interface (API; for example, supporting new input formats often requires modifying the source code), and its configuration interface was mainly designed for the convenience of the tests my colleagues and I have been conducting. However, since {\sc Effortless} is supposed to support use cases other than weak lensing cosmology as well, higher software design requirements need to be met.

\begin{table*}[t!]
    \centering
    \caption{Layout of {\sc Effortless} core modules.
    \label{tab:struct}}
    \begin{tabular}{ccll}
    \hline
        Module & Class & Description & Rough mapping to {\sc PyImcom} \\
    \hline
        \multirow{2}{*}{\tt io\_general} & {\tt InSlice} & Input image slice & Part of {\tt InImage} and {\tt InStamp} \\
        & {\tt OutSlice} & Output image slice & {\tt Block} \\
    \hline
        \multirow{2}{*}{\tt psfutil} & {\tt PSFModel} & Base class for PSF models & Part of {\tt InImage}, {\tt OutPSF} and {\tt \_LAKernel} \\
        & {\tt SubSlice} & Subslice of the output slice & {\tt OutStamp} \\
    \hline
        \multirow{2}{*}{\tt analysis} & {\tt MockImage} & --- & Wrapper of {\tt StarsAnal} \\
        & {\tt StarsCal} & Catalog of injected stars & Part of {\tt OutImage} \\
    \hline
    \end{tabular}
\end{table*}

Table~\ref{tab:struct} presents the OOP framework of {\sc Effortless}. Compared to that of {\sc PyImcom} \citep[see Table~1 of][]{2025ApJS..277...55C}, not only is it much simpler (which is part of the reason it is called ``effortless''), it is also more convenient: Users can easily inherit {\sc Effortless} core classes and overload their I/O methods to suit their needs. A module not shown in Table~\ref{tab:struct}, {\tt io\_pyimcom}, serves dual purposes: On the one hand, it harnesses {\sc PyImcom} utilities as much as possible (to avoid rewriting or duplicating the code); on the other hand, it provides an example of how to make use of the {\sc Effortless} API.

To achieve versatility, {\sc Effortless} also has a more general and flexible data model. It conceptualizes both input ({\tt InSlice}) and output ({\tt OutSlice}) images as regular arrays of pixels with WCSes attached to them and makes minimal additional assumptions. The classes are called ``slices'' because they are typically segments of larger arrays: For input images, only segments relevant to the region being processed (see Figure~\ref{fig:distort_mat} for the relative sizes of input images and the block) need to be stored in the memory; for output images, each block is a segment of a huge image \citep[``mosaic''; see Figure~4 of][]{2024MNRAS.528.2533H} that must be divided to limit the size of individual output files.

The workflow of {\sc Effortless} is simple: To reconstruct an {\tt OutSlice}, one identifies candidate {\tt InSlice}s (usually by searching in a catalog) with {\tt PSFModel}s attached to them, and loops over the {\tt SubSlice}s. To avoid clutter and make use of just-in-time compilation \citep{2015llvm.confE...1L}, several utility functions are wrapped into the {\tt routine} module; see the docstrings for detailed explanations.\footnote{\url{https://github.com/Roman-HLIS-Cosmology-PIT/effortless/blob/main/effortless/routine.py}} It is worth noting that each {\tt SubSlice} is equivalent to a group of $2 \times 2$ {\tt OutStamp}s in {\sc PyImcom}; both correspond to a PSF sampling point, i.e., where PSFs and related quantities are computed.

Subtleties of the {\sc Effortless} algorithm are configured via hyperparameters, which are the topic of the next section.

\subsection{Hyperparameters} \label{ss:params}

In each run, {\sc PyImcom} reads a JSON configuration file, converts it to a {\tt Config} instance, allows the user to modify its instance attributes, follows it to reconstruct a block, and finally writes (the actually used version of) it to the output FITS file. By contrast, the current version of {\sc Effortless} does not have a unified configuration interface; instead, hyperparameters only appear as class attributes of the classes. The underlying design philosophy is to make the mapping between hyperparameters and parts of the program more transparent; users are supposed to customize them in Python scripts and Jupyter notebooks. The apparent drawback is the lack of enforced bookkeeping, which is important for reproducibility; this needs to be addressed in the future.

\begin{table*}
    \centering
    \caption{{\sc Effortless} essential hyperparameters. The asterisk * denotes ``hard'' parameters that are determined by the input data.
    \label{tab:params}}
    \begin{tabular}{ccll}
    \hline
        Class & Symbol & Description & Rough mapping to {\sc PyImcom} or reference \\
    \hline
        \multirow{3}{*}{\tt InSlice} & {\tt NSIDE}* & Original input slice size in pixels & {\tt Settings.sca\_nside} \\
        & {\tt NLAYER}* & Number of input layers & Length of {\tt EXTRAINPUT} $+1$ \\
        & {\tt NOMASK} & Whether to ignore the input pixel mask & Section~\ref{sec:calibr} of this paper \\
    \hline
        \multirow{5}{*}{\tt OutSlice} & {\tt NSUB} & Output slice size in subslices & First component of {\tt OUTSIZE} $/2$ \\
        & {\tt NPIX\_SUB} & Subslice size in pixels & Second component of {\tt OUTSIZE} $\times 2$ \\
        & {\tt NPIX\_TOT} & Output slice size in pixels & Product of first two components of {\tt OUTSIZE} \\
        & {\tt SIGMA} & Target output PSF width in native pixels & {\tt EXTRASMOOTH} (for Gaussian target output PSFs)\\
        & {\tt SAVE\_ALL} & Whether to save individual regridded images & \citet{2026arXiv260706674C} \\
    \hline
        \multirow{6}{*}{\tt PSFModel} & {\tt NPIX}* & PSF array size in native pixels & --- (Note: This is different from {\tt NPIXPSF}.) \\
        & {\tt SAMP}* & Oversampling rate of PSF arrays & Last component of {\tt INPSF} \\
        & {\tt NTOT}* & PSF array size in oversampled pixels & --- \\
        & {\tt YXCTR}* & PSF array center in oversampled pixels & --- \\
        & {\tt BL\_CIRC} & Circular bandlimit in Fourier space & Section~\ref{ss:limit} of this paper \\
        & {\tt BL\_INNER} & Inner bandlimit for fixing annular artifacts & Section~\ref{ss:limit} of this paper  \\
    \hline
        \multirow{5}{*}{\tt SubSlice} & {\tt ACCEPT} & Acceptance radius for selecting input pixels & {\tt INPAD}; Section~\ref{ss:accept} of this paper \\
        & {\tt REJECT} & Rejection radius for masking output pixels & Section~\ref{sec:mask} of this paper \\
        & {\tt MASK\_THR} & Threshold for number of masked input pixels & Section~\ref{sec:mask} of this paper \\
        & {\tt NDIFF} & Number of iterations for weight diffusion & Section~\ref{sec:mask} of this paper \\
        & {\tt RENORM} & Whether to renormalize weights & Section~\ref{sec:mask} of this paper \\
    \hline
        \multirow{2}{*}{\tt StarsCal} & {\tt DISTTHR} & Distance threshold for selecting stars & Section~\ref{sec:mask} of this paper \\
        & {\tt KMAX} & Maximum order of subpixel features & Section~\ref{sec:calibr} of this paper \\
    \hline
    \end{tabular}
\end{table*}

{\sc Effortless} hyperparameters are listed in Table~\ref{tab:params}. Those with references to later sections of this paper will be defined and explained therein; the rest of this section discusses those corresponding to {\sc PyImcom} configuration entries. In general, simulations in this series of papers are configured following \citet{2026ApJ...998..304C}; see Section~2.2 of \citet{2026arXiv260706674C} for some further explanations.

{\tt InSlice} class attributes.
\begin{itemize}
    \item ${\tt NSIDE} = 4088$: This is always the case for Roman images, since each Roman H4RG-10 detector \citep{2020JATIS...6d6001M} has $4088^2$ science pixels.
    \item ${\tt NLAYER} = 4$: Layers are different versions of the input images to be reconstructed; since the weights $T_{{\boldsymbol \alpha} {\boldsymbol i}}$ in Equation~(\ref{eq:transform}) do not depend on the signals $I_{\boldsymbol i}$, they can be reused. Specifically, the four input layers in \citet{2026arXiv260706674C} and this paper are: simulated science images \citep[{\tt \textquotesingle SCI\textquotesingle};][]{2025MNRAS.544.3799O}, injected stars drawn by {\sc GalSim} \citep[{\tt \textquotesingle gsstar14\textquotesingle};][]{2015A&C....10..121R}, and simulated white ({\tt \textquotesingle whitenoise10\textquotesingle}) and $1/f$ ({\tt \textquotesingle 1fnoise9\textquotesingle}) noise frames \citep{2024MNRAS.528.2533H}. The analysis in this work focuses on the injected stars.
\end{itemize}

{\tt OutSlice} class attributes.
\begin{itemize}
    \item ${\tt NSUB} = 42$, ${\tt NPIX\_SUB} = 64$, ${\tt NPIX\_TOT} = 2688$: In \citet{2026ApJ...998..304C}, each block has $84^2$ postage stamps (with two rows or columns of padding regions on each side), and each postage stamp has $32^2$ output pixels.
    \item {\tt SIGMA}: This is just the target output PSF width $\sigma$ in Equations~(\ref{eq:gauss}) and (\ref{eq:gauss_in}). Following the benchmark cases in \citet{2026ApJ...998..304C}, the FWHMs of the target output PSFs in the Y106, J129, H158, and F184 bands are $2.0$, $2.1$, $2.2$, and $2.3$ native pixels, respectively.
    \item ${\tt SAVE\_ALL} = {\tt True}$: As illustrated in \citet{2026arXiv260706674C}, {\sc Effortless} has the capability of reconstructing individual input images. Throughout this work, measurements are performed on reconstructed individual images, hence they need to be saved separately.
\end{itemize}
It is worth emphasizing that, to only reconstruct a relatively small piece of the sky (e.g., for studying a nearby galaxy), one simply needs to use a much smaller {\tt NPIX\_SUB}. (To increase the resolution of output images, one uses an output WCS with small {\tt CDELT} values to initialize the {\tt OutSlice} instance.)

{\tt PSFModel} and {\tt SubSlice} class attributes.
\begin{itemize}
    \item ${\tt NPIX} = 32$, ${\tt SAMP} = 8$, ${\tt NTOT} = 256$, ${\tt YXCTR} = 127.5$: These are ``hard'' parameters describing the input PSF format. The center of a PSF image can be either at the center of a central pixel or between the four central pixels (or anywhere, in principle); the input PSFs used in this series of work follow the latter convention. The ${\tt YXCTR}$ needs to be specified so that the target output PSFs are centered in exactly the same way to avoid aliasing in Equation~(\ref{eq:T_solution}).
    \item ${\tt ACCEPT} = 8$: This is the ``acceptance radius'' use to select input pixels (the term was adopted from the {\sc Imcom} terminology). For input pixel windows with different shapes, it can have different meanings \citep[see Equations~(8) and (17) of][]{2025ApJS..277...55C}. By default, {\sc Effortless} uses square input pixel windows of size $(2 \times {\tt ACCEPT})^2$ native pixels (see the end of Section~\ref{ss:accept} for some discussion). Note that the {\tt INPAD} configuration entry of {\sc PyImcom} is expressed in arcseconds; in benchmark cases of the Cholesky linear algebra kernel in \citet{2026ApJ...998..304C}, ${\tt INPAD} = 1.24 \,{\rm arcsec} > 0.88 \,{\rm arcsec} = 8$ native pixels.
\end{itemize}

\section{Understanding PSF Residuals} \label{sec:residual}

As outlined in Section~\ref{ss:math_gen}, there are several imperfections involved in the image reconstruction process. In this section, I examine three sources of PSF residuals ($\Psi'_{\boldsymbol \alpha} - \Gamma'$), circular bandlimits (Section~\ref{ss:limit}), finite sampling (Section~\ref{ss:samp}), and finiteness of input pixel windows (Section~\ref{ss:accept}). Throughout this section, all selected input pixels are assumed to be available; the unavailability of some input pixels is addressed in Section~\ref{sec:mask}. Furthermore, all distortion matrices Equation~(\ref{eq:distort}) are assumed to be the identity matrix, which is reasonable given what is shown in Section~\ref{ss:distort} (when $\kappa, |{\boldsymbol \gamma}| \ll 1$, $\varphi$ does not matter much for Equation~(\ref{eq:gauss_in})).

\subsection{Circular Bandlimit} \label{ss:limit}

\begin{figure*}
    \centering
    \includegraphics[width=\textwidth]{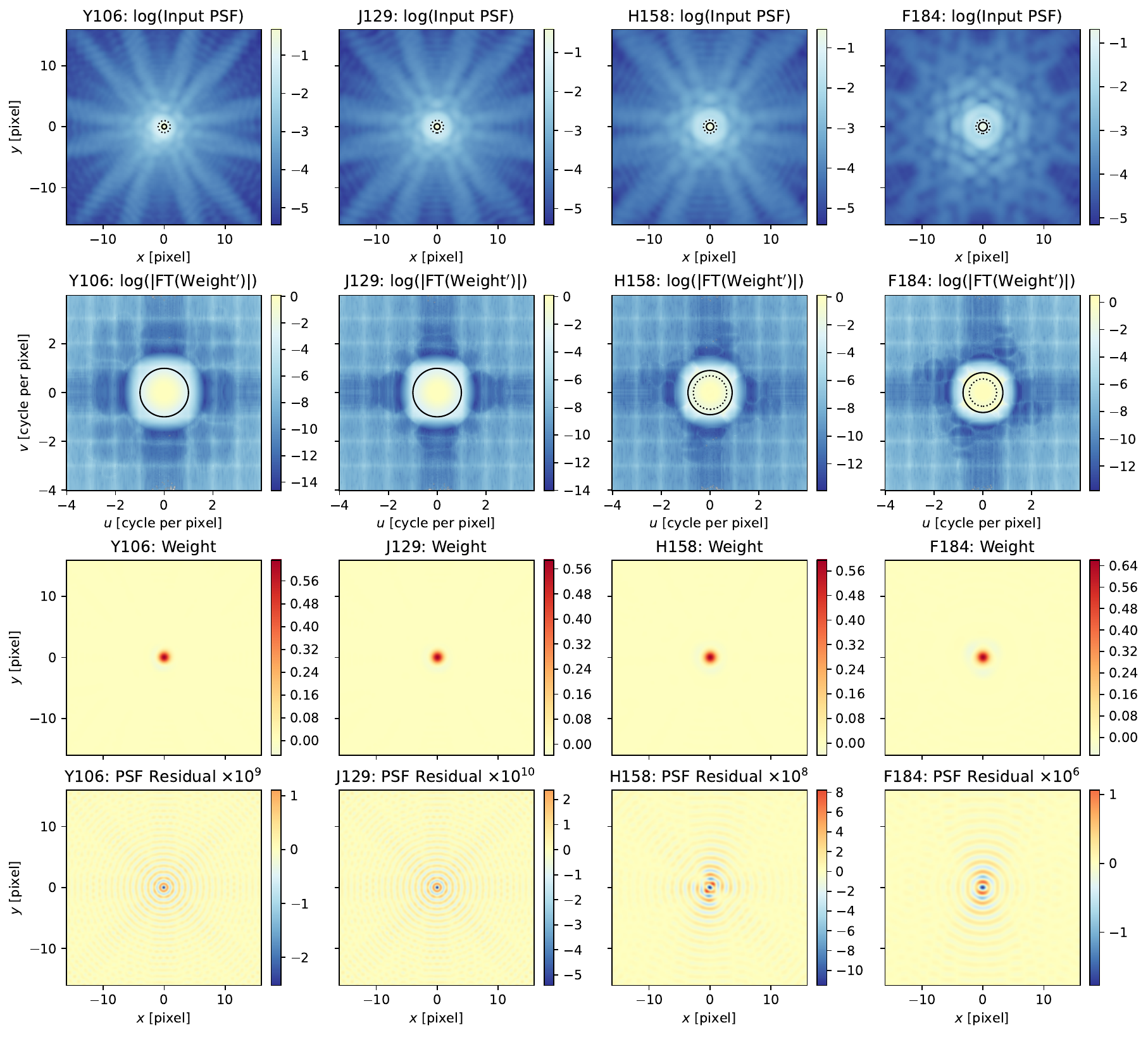}
    \caption{Summary of how the weight fields are computed. From left to right, the four columns correspond to the Y106, J129, H158, and F184 bands, respectively. The first row shows example PSFs in different bands in logarithmic scale, so that structures in the outer regions can be easily seen. In the inner regions, half widths at half maxima (HWHMs) of the input PSFs (mainly characterized by the wavelength-to-aperture diameter ratios $\lambda/D$) and target output PSFs (chosen to be bivariate Gaussian functions) are displayed as solid and dotted black circles, respectively. The second row shows (the magnitudes of) the quotients of the target output PSFs and the pixelated input PSFs in Fourier space, again in logarithmic scale. The artifacts necessitate circular bandlimits, which are shown as solid black circles. In the H158 and F184 bands, inner bandlimits are used to further reduce artifacts; these are shown as dotted black circles. The third row shows the resulting weight fields, normalized to ensure the conservation of surface brightness when the sampling rate is once per pixel. Even in the absence of effects caused by finite sampling and finiteness of input pixel windows, the PSF residuals are very small but not zero due to numerical artifacts; these are shown in the last row, multiplied by powers of $10$ to avoid clutter.
    \label{fig:limit_all}}
\end{figure*}

Figure~\ref{fig:limit_all} illustrates several key ingredients in Section~\ref{ss:math_gen}. The first row shows example (forward) input PSFs $G$ in different bands. The inner regions are dominated by Airy disks characterized by the wavelength-to-aperture diameter ratios $\lambda/D$, hence the widths (related to solid black circles) are larger in redder bands, as expected; the benchmark (backward) target output PSFs $\Gamma'$ (related to dotted black circles) in \citet{2026ApJ...998..304C} are indeed slightly wider than the corresponding input PSFs. The outer regions are mainly diffraction spikes, with more complicated structures in the F184 band; despite the least severe undersampling, such structures cause additional difficulty in the F184 band (see below). The second row shows the quotients ${\tilde T} = {\tilde \Gamma}' / {\tilde G}'$ in Equation~(\ref{eq:T_solution}). Because of pixelation, the amplitudes of ${\tilde T}$ are practically infinite when both wavenumbers are non-zero integers (in cycles per pixel); in Figure~\ref{fig:limit_all}, these are fixed by taking geometric averages of their nearest neighbors, but the ``grid lines'' are still very easy to see. Therefore, {\sc Effortless} applies some bandlimit in each band to avoid these numerical artifacts; since PSFs are approximately symmetric, such bandlimit is circular. As a hyperparameter, {\tt BL\_CIRC} in Table~\ref{tab:params} is in units of ${\tt NPIX}^{-1} = 1/32$ cycles per pixel to facilitate operations on discrete arrays; specific values in different bands are chosen in Section~\ref{ss:samp}.

The third row shows the resulting weight fields $T$. In each band, the weight field has a central positive ``bulb,'' surrounded by a zero annulus, which is in turn surrounded by a negative annulus. Since the target output PSF $\Gamma'$ is a Gaussian, were the input PSF $G'$ also a Gaussian, we would expect $T$ to be another Gaussian; such annular structures are caused by the non-Gaussianity of the input PSF. The last row shows PSF residuals purely due to the bandlimits (i.e., without finite sampling effects), $T * G' - \Gamma'$ (here $*$ denotes convolution). Since the bandlimits are circular, these PSF residuals are alternating positive and negative rings, which are finer in bluer bands with larger {\tt BL\_CIRC} and coarser in redder bands with smaller {\tt BL\_CIRC}. Such PSF residuals are mostly negligible because of their small magnitudes, at the $10^{-9}$, $10^{-10}$, and $10^{-8}$ levels in the Y106, J129, and H158 bands, respectively; the PSF residual in the F184 band is relatively large, at the $10^{-6}$ level, due to the additional difficulty mentioned above and addressed below. In general, PSF residuals purely due to circular bandlimits are small compared to those caused by other effects.

\begin{figure*}
    \centering
    \includegraphics[width=\textwidth]{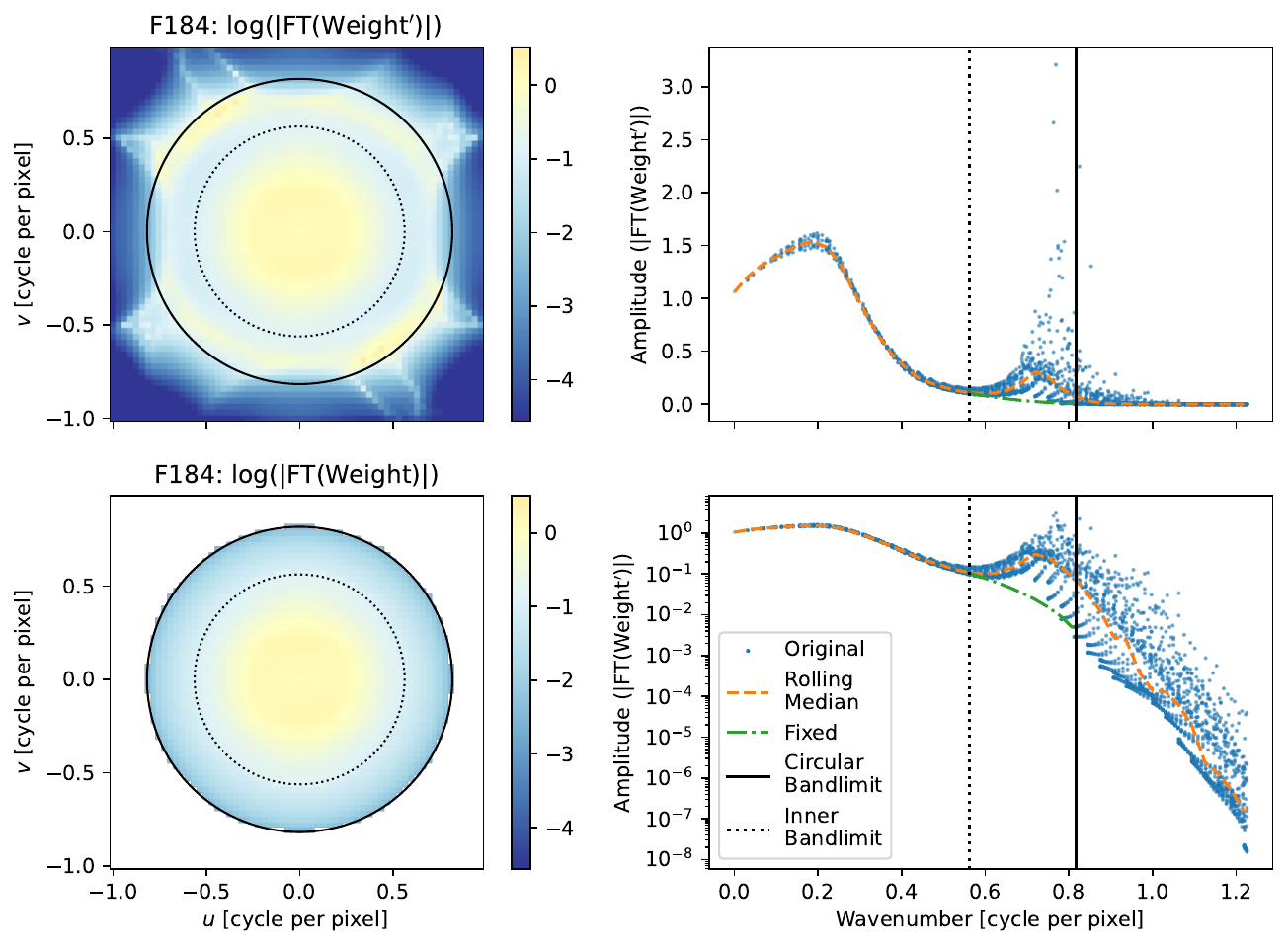}
    \caption{Application of circular bandlimits. The upper left panel shows the central part of the last panel of the third row of Figure~\ref{fig:limit_all}, i.e., quotient of the target output PSF and the pixelated input PSF in the F184 band in Fourier space. Note that the minimum value is different so that inner structures can be better seen. The solid and dotted black circles are the circular and inner bandlimits, respectively. The lower left panel shows the same quotient, but with artifacts removed by applying the bandlimits; the weight field is obtained by performing inverse Fourier transform on it. The right column explains the application of the inner bandlimit by plotting the amplitudes of the quotient versus the wavenumbers of the modes as blue data points. The $y$-axes of the upper right and lower right panels are in linear and logarithmic scales to better display inner and outer features, respectively; the ingredients are the same. Rolling medians of the blue data points are shown as orange dashed curves; between the inner and circular bandlimits (dotted and solid black vertical lines), the modified amplitudes are shown as green dash-dotted curves.
    \label{fig:limit_F184}}
\end{figure*}

Figure~\ref{fig:limit_F184} addresses the additional difficulty in the F184 band. The upper left panel shows the central part of ${\tilde T} = {\tilde \Gamma}' / {\tilde G}'$. Evidently, it has an annulus of significant magnitudes (larger than modes in the core) inside the chosen circular bandlimit. Such annulus is likely an artifact caused by the complicated structures seen in the upper right panel of Figure~\ref{fig:limit_all} and needs to be removed. However, if we simply apply a small circular bandlimit, the resulting PSF leakage is large. Therefore, {\sc Effortless} allows users to set an inner bandlimit; it fits a third-order polynomial (amplitude as a function of wavenumber) to the edge of the inner region while enforcing zero outside the ``outer'' region, and then uses it to rescale modes between this inner bandlimit (dotted black circle) and the ``outer'' circular bandlimit (solid black circle). Such fitting and rescaling process is illustrated in the right column of Figure~\ref{fig:limit_F184}. While this figure only illustrates this special handling in the F184 band, additional testing indicates that it is also beneficial in the H158 band. The hyperparameter {\tt BL\_INNER} is also in units of ${\tt NPIX}^{-1} = 1/32$ cycles per pixel; it is chosen to be $18$ in the F184 band and $22$ in the H158 band for simulations in this series of papers.

\subsection{Finite Sampling} \label{ss:samp}

Finite sampling effects have been studied in \citet{2026AJ....171..140C}. According to 1D experiments in that paper, PSF residuals due to finite sampling are simple wave packets; when the output pixel is misaligned with the input pixel array by $\Delta x$ (in units of native pixels), the profile of the wave packet is unaffected, and its phase in Fourier space is multiplied by $e^{-2\pi i \Delta x}$. Therefore, when combining two 1D images separated by half a native pixel, the PSF residuals can exactly cancel (see Figure~8 of that paper). In 2D and with more realistic input PSFs, the pattern is similar, but with some nuances.

\begin{figure*}
    \centering
    \includegraphics[width=\textwidth]{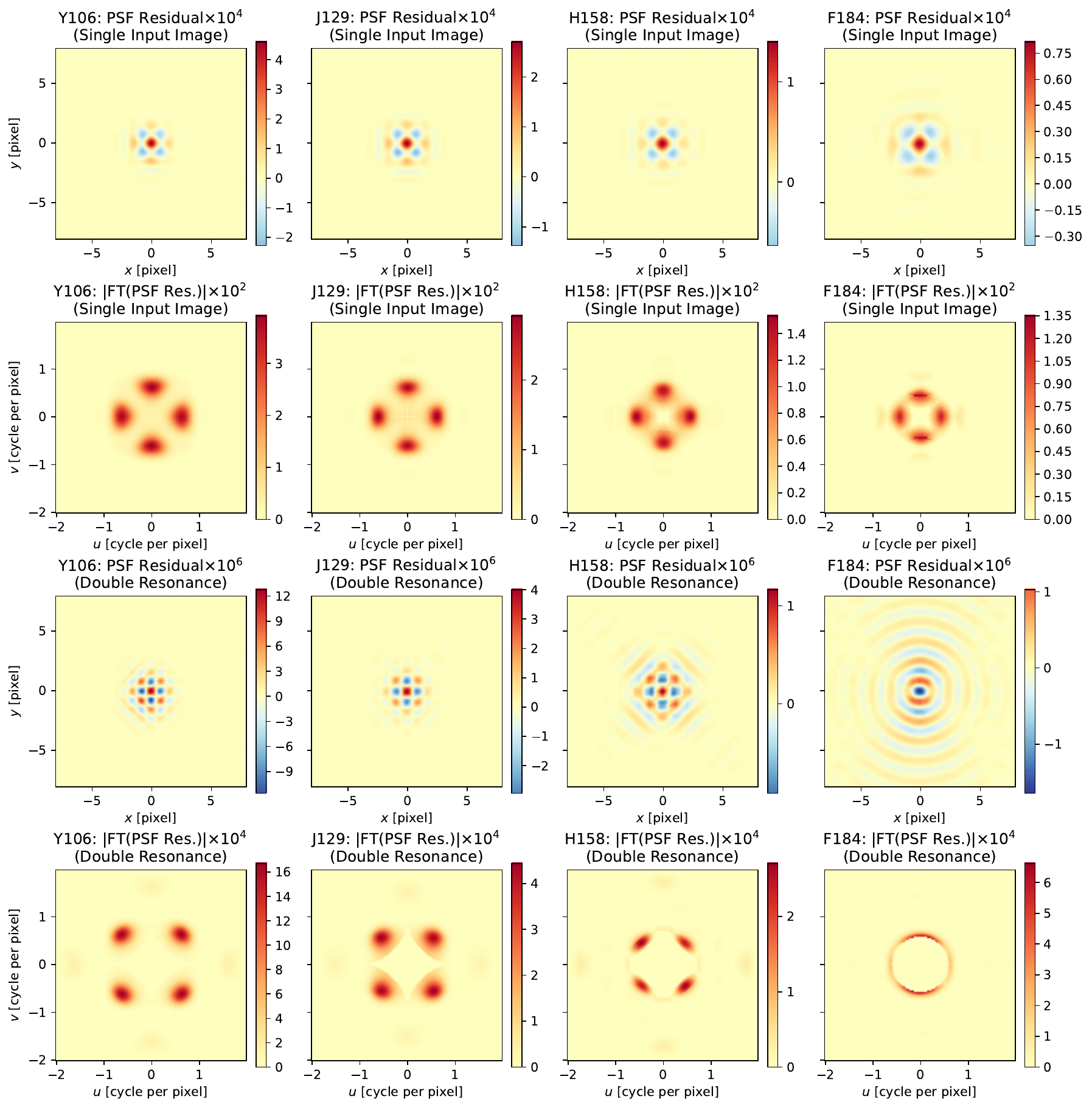}
    \caption{PSF residuals caused by finite sampling. From left to right, the four columns correspond to the Y106, J129, H158, and F184 bands, respectively. The first and third rows present the PSF residuals in real space, while the second and fourth rows present them in Fourier space. In each row, the PSF residuals are multiplied by some power of $10$ to avoid clutter. Throughout this figure, the PSF residuals are mainly caused by finite sampling (those shown in the last row of Figure~\ref{fig:limit_all} are negligible here); the first two rows present PSF residuals for single input images, while the last two rows present PSF residuals for the ``double resonance'' dithering pattern (two images misaligned by half a pixel in each direction).
    \label{fig:samp_all}}
\end{figure*}

Figure~\ref{fig:samp_all} presents example PSF residuals caused by finite sampling. The first two rows show PSF residuals for single images, whose subpixel positions are all $(\Delta x, \Delta y)^{\rm T} = (0, 0)^{\rm T}$ (still in native pixels). The wave packets are similar to those seen in Figure~5 of \citet{2026AJ....171..140C}, with narrower profiles and lower oscillation frequencies. Such differences are mainly due to the fact that I was using a wider (by a factor of $1.5$) target output PSF in that paper; the impact of target output PSF width will be systematically explored in a future paper. Given the input and target output PSFs in this work (see the first row of Figure~\ref{fig:limit_all}), in redder bands, the profiles are wider in real space, the corresponding mode groups are more localized in Fourier space; the oscillation frequencies are lower, and the magnitudes are smaller. When the subpixel positions are all $(\Delta x, \Delta y)^{\rm T} = (1/2, 1/2)^{\rm T}$ (not shown in this figure), the PSF residuals are basically the negatives of those when $(\Delta x, \Delta y)^{\rm T} = (0, 0)^{\rm T}$.

The combination of $(0, 0)^{\rm T}$ and $(1/2, 1/2)^{\rm T}$ constitutes the ``double resonance'' ideal dithering pattern as defined in Section~5.4 of \citet{2026AJ....171..140C}; extrapolated from the 1D experiments therein, such a combination should have basically zero PSF residuals. In reality, as shown in the last two rows of Figure~\ref{fig:samp_all}, we see that the ``double resonance'' pattern does substantially reduce PSF residuals, from the $10^{-4}$ level to the $10^{-6}$ level in real space and from the $10^{-2}$ level to the $10^{-4}$ level in Fourier space. In the F184 band, this happens to be the level of PSF residuals caused by circular bandlimits even in the absence of finite sampling effects, hence the third panel of the last column of Figure~\ref{fig:samp_all} is very similar to the bottom right panel of Figure~\ref{fig:limit_all}. In the H158 band, we also see alternating positive and negative rings (and corresponding features in Fourier space), but they are much fainter; in the J129 and Y106 bands, such structures are not visible. In the three bluer bands, PSF residuals of the ``double resonance'' pattern are wave packets with different directions and oscillation frequencies; these can be further canceled out in the $2 \times 2$ ideal dithering pattern (all four combinations with $\Delta x, \Delta y \in \{0, 1/2\}$; not shown in Figure~\ref{fig:samp_all}).

\begin{figure}
    \centering
    \includegraphics[width=\columnwidth]{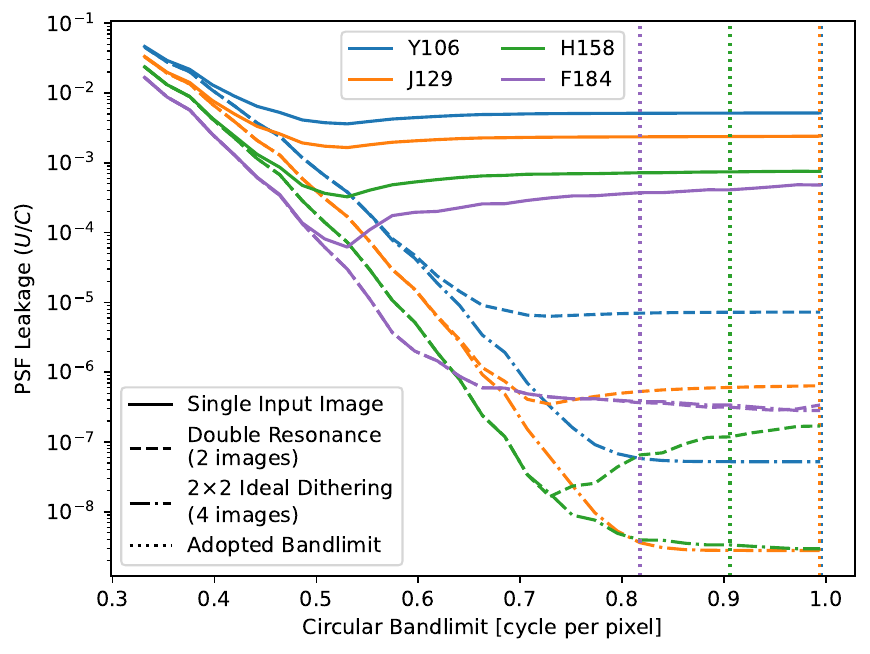}
    \caption{Tuning circular bandlimits given finite sampling. This figure plots the PSF leakage ($U/C$) versus the circular bandlimit (with the inner bandlimits kept to their default values). Results for the Y106, J129, H158, and F184 bands are shown in blue, orange, green, and purple, respectively. The solid, dashed, and dot-dashed curves are results for single input images, the ``double resonance'' dithering pattern, and the $2 \times 2$ ideal dithering pattern (four images aligned or misaligned by half a pixel in each direction). Circular bandlimits adopted in this work are shown as dotted vertical lines.
    \label{fig:samp_limit}}
\end{figure}

The main lesson here is that, when reconstructing single input images, the PSF residuals caused by finite sampling are relatively large (i.e., exceeding weak lensing requirements), but they mostly follow simple patterns that can be either canceled out by combining a group of properly dithered images (which is not necessarily available) or calibrated out by leveraging the knowledge about wave packets (see Section~\ref{sec:calibr}). Therefore, when one sets circular bandlimits (see Section~\ref{ss:limit}), one should keep such potential in mind. For simulations in this series of papers, bandlimits are tuned in Figure~\ref{fig:samp_limit}. For single input images (solid curves), PSF leakages first decrease with larger circular bandlimits, reach their minima slightly after $0.5$ cycles per pixel, and then increase before converging to certain values. However, if we choose those minimum points, we lose the opportunity to obtain better measurements if effects of wave packets can be largely eliminated, as evidenced by PSF leakages of the double resonance (dashed curves) and $2 \times 2$ (dash-dotted curves) ideal dithering patterns. Considering PSF leakages in different scenarios and numerical artifacts at non-zero integer wavenumbers (see the second row of Figure~\ref{fig:limit_all}), I chose {\tt BL\_CIRC} values (in units of ${\tt NPIX}^{-1} = 1/32$ cycles per pixel) of $18.5\sqrt{2}$ for the F184 band, $20.5\sqrt{2}$ for the H158 band, and $22.5\sqrt{2}$ for the J129 and Y106 bands\footnote{The choice of half-integer multiplied by $\sqrt{2}$ was made for historical reasons during the development of {\sc Effortless} {that are no longer relevant}. Since the curves in Figure~\ref{fig:samp_limit} are smooth, details of the specific values should not matter.} for simulations in this work, as shown in Figure~\ref{fig:samp_limit} as dotted vertical lines.

\subsection{Input Pixel Window} \label{ss:accept}

As mentioned in Section~\ref{ss:math_gen}, practical image reconstruction algorithms only assign weights to adjacent input pixels. For {\sc Effortless}, ``adjacent'' means within square input pixel windows of size $(2 \times {\tt ACCEPT})^2$ native pixels, where {\tt ACCEPT} is the acceptance ``radius'' hyperparameter, chosen to be $8$ for simulations in this series of papers.

\begin{figure*}
    \centering
    \includegraphics[width=\textwidth]{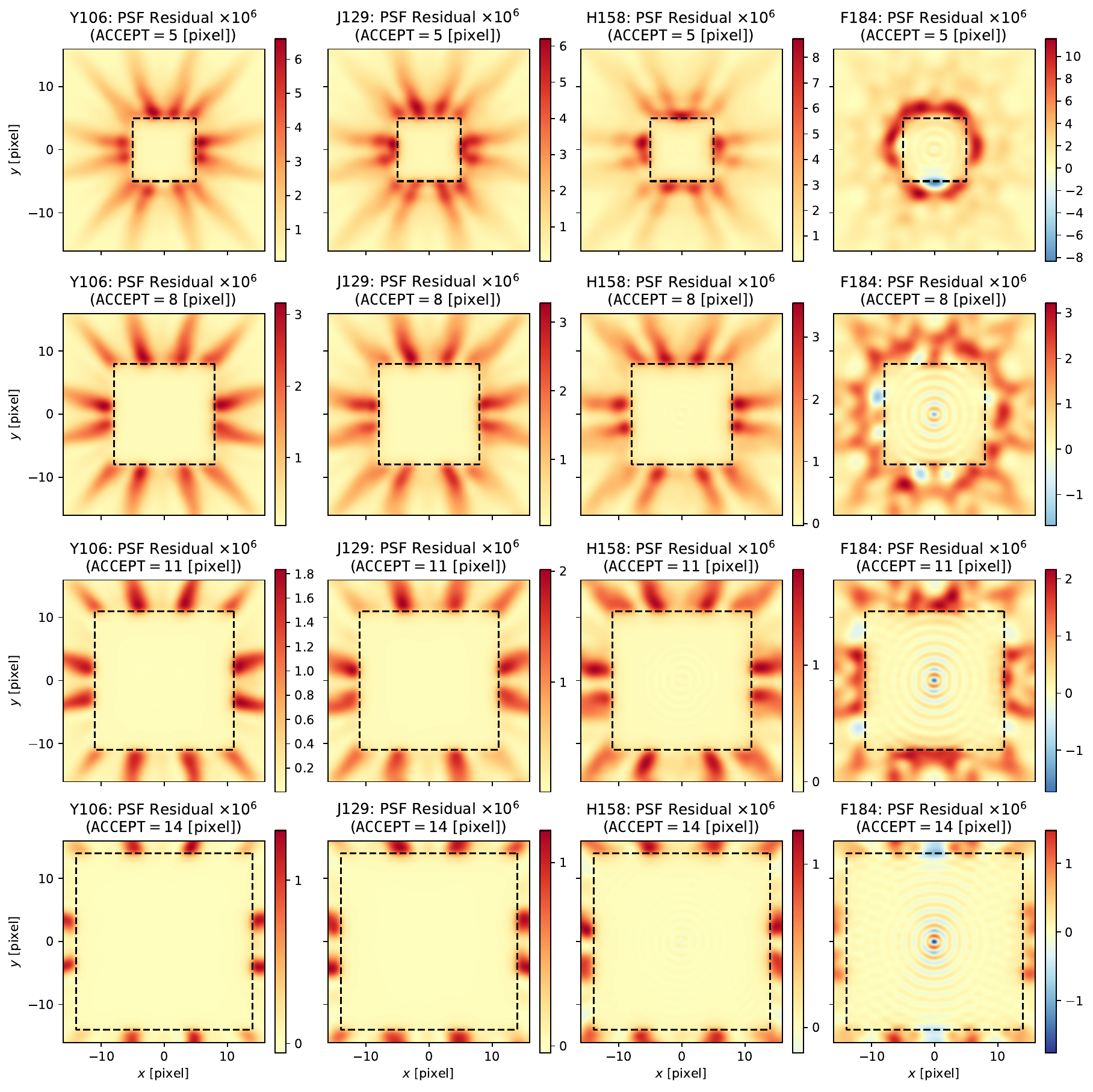}
    \caption{PSF residuals caused by the finiteness of input pixel windows. To isolate this causal relationship, effects due to finite sampling (see Section~\ref{ss:samp}) are not included here. From left to right, the four columns correspond to the Y106, J129, H158, and F184 bands, respectively. From top to bottom, the four rows correspond to acceptance ``radii'' of $5$, $8$, $11$, and $14$ native pixels, respectively. The corresponding input pixel windows are displayed as dashed black squares. All the panels visualize the PSF residuals in linear scale (with different extreme values though) and multiply them by a factor of $10^6$ to avoid clutter.
    \label{fig:accept_all}}
\end{figure*}

The finiteness of input pixel windows amount to a truncation of the weight fields $T$. Since actual input PSFs are larger than a patch of $16^2$ native pixels, such truncation causes additional PSF residuals, as expected. To isolate the truncation effects, Figure~\ref{fig:accept_all} shows PSF residuals with finite acceptance ``radii'' but without finite sampling effects (see Section~\ref{ss:samp}). Analogous to what was found in Appendix of \citet{2026AJ....171..140C}, truncation means outside regions corresponding to the input pixel windows, PSF residuals are ``blurred'' versions of the outer regions of the input PSFs: In the Y106, J129, and H158 bands, we mostly see the diffraction spikes; in the F184 band, we see the ``complicated structures'' in addition (and the alternate positive and negative rings due to circular bandlimits in the inner regions as well). This effect can be understood as follows: When {\sc Effortless} tries to reconstruct outer regions of a target output PSF $\Gamma'$, in principle it should assign negative weights (probably with small absolute values) to central regions of the input PSF $G'$, so that positive values in its outer regions (note that input PSFs are non-negative) are offset; however, because of the input pixel window, {\sc Effortless} actually assigns zero weights to the central regions of $G'$, so that outer regions of the resulting output PSFs $\Psi'_{\boldsymbol \alpha}$ are outer regions of $G'$ convolved with the weight field $T$.\footnote{By the same argument, the ``postage stamp boundary effects'' discussed in \citet{2025ApJS..277...55C} are also PSF residuals due to finiteness of input pixel windows.} Therefore, such PSF residuals cannot be removed without increasing {\tt ACCEPT}.

\begin{figure}
    \centering
    \includegraphics[width=\columnwidth]{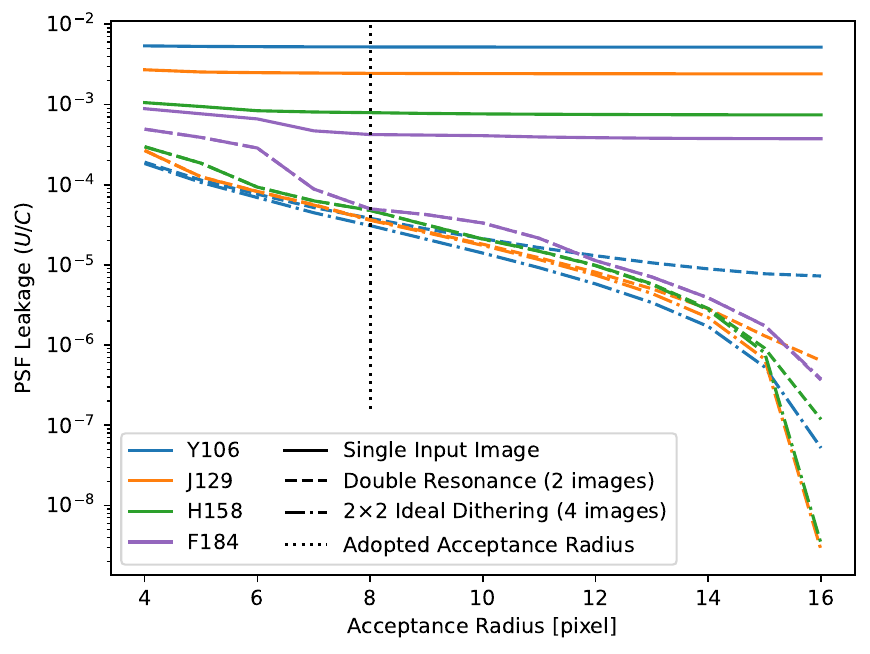}
    \caption{Tuning acceptance ``radius'' given finite sampling. This figure plots the PSF leakage ($U/C$) versus the acceptance ``radius'' (with the circular and inner bandlimits kept to their default values). The coloring scheme and the line styles are the same as in Figure~\ref{fig:samp_limit}, except that here the black dotted vertical line denotes the acceptance ``radius'' adopted in this work.
    \label{fig:accept_tuning}}
\end{figure}

While diffraction spikes of bright objects are worth removing \citep[see Section~3 of][for a fuller discussion]{2026arXiv260706674C}, for faint sources the PSF residuals due to the finiteness of input pixel windows may or may not matter, as they are only at the $10^{-6}$ level. As shown in Figure~\ref{fig:accept_all}, while a larger acceptance radius {\tt ACCEPT} can reduce the size of non-zero regions of the PSF residuals, it barely reduces their magnitude; meanwhile, the computational costs scale as ${\tt ACCEPT}^2$ (reconstruction weights need to be computed for selected input pixels), hence a balance needs to be found between these two factors. To inform this decision, Figure~\ref{fig:accept_tuning} plots PSF leakages versus different acceptance radii, this time with finite sampling effects. For single input images, the PSF residuals are dominated by the wave packets, and the outer regions do not matter; however, to retain the potential of calibrating finite sampling effects, the acceptance radius should not be too small. For simulations in this series of papers, {\tt ACCEPT} is chosen to be $8$ in all bands, as indicated by the black dotted vertical line. Note that the sharp drop at ${\tt ACCEPT} = 16$ native pixels in some cases is because input PSFs used in this work only span $32^2$ native pixels (${\tt NPIX} = 32$, see Section~\ref{ss:params}) and should not be relied upon for real-world PSFs.

Before concluding this section on input pixel windows, there are two further considerations related to the input pixel windows to keep in mind. First, since the hyperparameter is called acceptance ``radius,'' not acceptance ``half side length,'' it is natural to imagine circular input pixel windows instead of square ones. During the development of {\sc Effortless}, the former was tried, and they indeed change the shape of boundaries between inner and outer regions of PSF residuals from square to circular. However, with the same {\tt ACCEPT}, this slightly increases the PSF leakages, and the reduction in computational costs is modest. Therefore, it was decided to keep using square input pixel windows; the term acceptance ``radius'' can be understood in the sense of the $L$-infinity norm. Second, the current version of {\sc Effortless} treats boundaries of input images in a conservative way: If an entire row or column in the array of $(2 \times {\tt ACCEPT})^2$ selected pixels is unavailable, it does not try to reconstruct the output signal at that position. Investigating other ways of handling detector boundaries is left for future work.

\section{Handling Input Pixel Masks} \label{sec:mask}

Throughout Section~\ref{ss:accept}, I have been assuming that all pixels within the square input pixel windows are available. However, the Roman WFI does have many inoperable pixels, either permanent or temporary; meanwhile, since Roman will be in space, the WFI will be more subject to cosmic ray hits than ground-based instruments. According to Table~2 of \citet{2024MNRAS.528.2533H}, $\sim 3\%$ of the native pixels are not advisable to use for weak lensing cosmology (and excluding $8$ rows or columns on each boundary adds $\sim 0.4\%$ to this quote). It would be a huge waste of data if {\sc Effortless} was only to reconstruct ``fortunate'' portions of the sky where all selected input pixels are available; that is why here is a section dedicated to the handling of input pixel masks.

\begin{figure*}
    \centering
    \includegraphics[width=\textwidth]{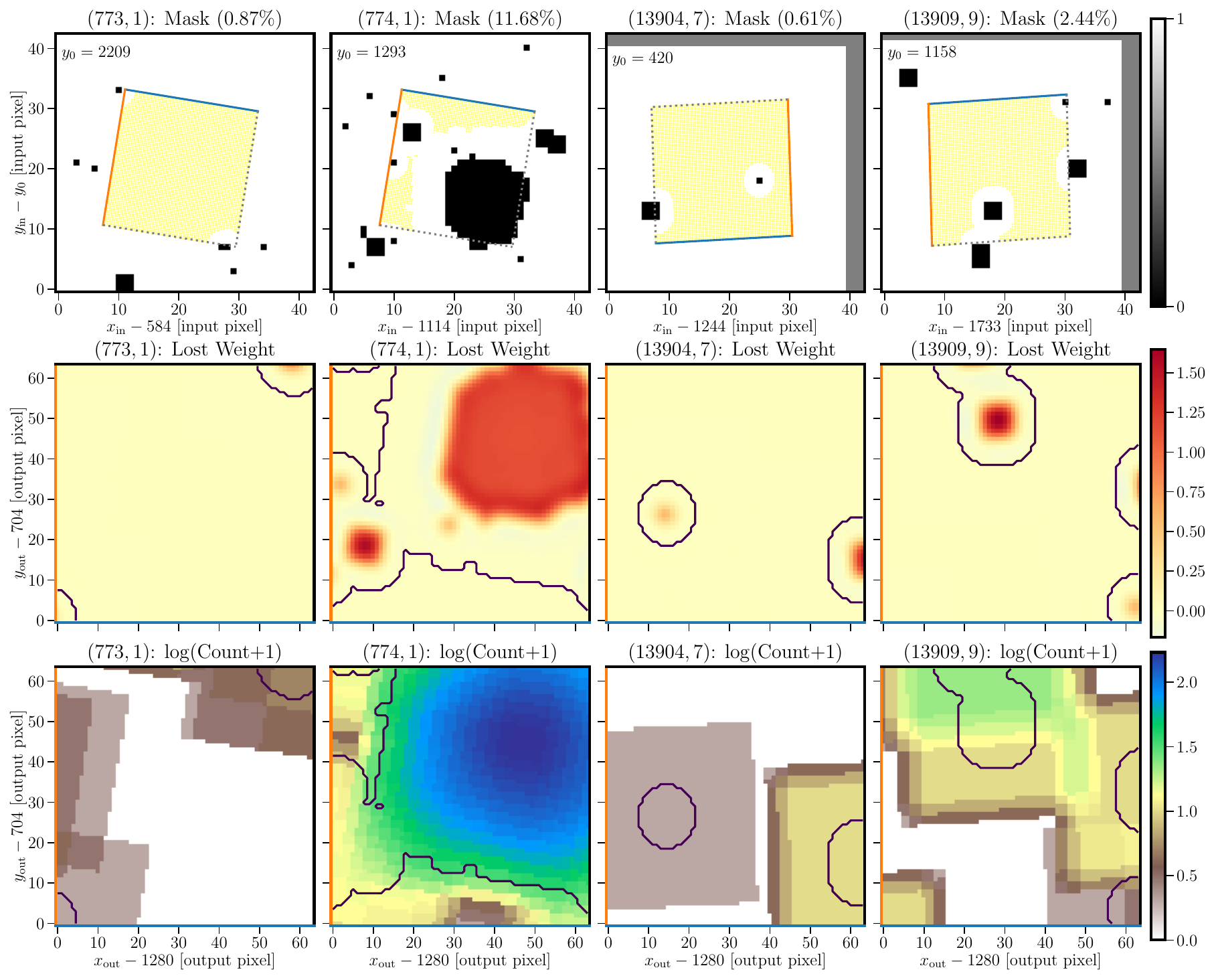}
    \caption{Four examples illustrating the propagation of pixel masks. The four columns correspond to the same four input images as in Figure~\ref{fig:distort_mat}; subslice $(20, 11)$ of block $(9, 20)$ is chosen for this figure to present a variety of scenarios. In each column, the upper panel displays the relevant part of an input pixel mask, with masked pixels shown in black and the fraction of masked pixels annotated on top of the panel; unmasked output pixels are shown in yellow, and the blue and orange line segments correspond to the $x$- and $y$-axes of the other two panels, respectively. The location of the ``relevant part'' is indicated by the $x_0$ value in the $x$-axis label and the $y_0$ value annotated in the upper left corner; for the last two images, the ``relevant parts'' are smaller, and the unused parts of the corresponding panels are colored gray (the masks are boolean). The second and third panels of each column visualize the total lost weights and numbers of lost neighboring input pixels of the output pixels, respectively; boundaries of masked regions are shown as purple curves. Note that all panels in the last two rows are showing the same region, subslice $(20, 11)$ of block $(9, 20)$, and thus they are in output pixel coordinates; panels in the first row are in input pixel coordinates.
    \label{fig:mask_examples}}
\end{figure*}

Figure~\ref{fig:mask_examples} illustrates the propagation of pixel masks in {\sc Effortless}. The four input images are the same as those in Figure~\ref{fig:distort_mat}, and the subslice is deliberately chosen to present a variety of scenarios. As shown in the first row, $(773, 1)$ only has a few masked input pixels near the boundaries of the subslice, while $(774, 1)$ is subject to heavy losses because of a large clump of masked input pixels; the situations of $(13904, 7)$ and $(13909, 9)$ are between those two extreme cases, with isolated or small groups of masked input pixels overlapping with the subslice. The second row shows the sum of weights $T_{{\boldsymbol \alpha} {\boldsymbol i}}$ masked input pixels would contribute to each output pixel. Corresponding to the inner structures of weight fields (see the third row of Figure~\ref{fig:limit_all}), the regions of masked input pixels have positive total lost weights, and surrounding them are zero bands and then negative bands. From the $(774, 1)$ panel, we also see that boundaries of the ``clump'' region have larger total lost weights than its inner regions; this is because in the inner regions negative reconstruction weights (see the third row of Figure~\ref{fig:limit_all}) partially cancel positive ones.

While it is feasible to set a threshold for the total lost weight, this would require computing all reconstruction weights and a lot of partial sums. {\sc Effortless} propagates input pixel masks to output pixel masks in a more economical manner. Specifically, it rejects output pixels with any masked input pixel within the user-specified rejection radius; unlike the acceptance ``radius'' (see Section~\ref{ss:accept}), the rejection radius does correspond to circular regions. The hyperparameter {\tt REJECT} is an integer in units of output pixels and is chosen to be $8$ (which corresponds to $2.84$ native pixels) for simulations in this work. Furthermore, {\sc Effortless} also masks output pixels with more than the user-specified number of masked neighboring input pixels. The hyperparameter {\tt MASK\_THR} is a dimensionless integer (number of pixels) and is chosen to be $32$; the associated ``neighborhood'' is defined as $16^2$ input pixels, which coincides with the input pixel window set by {\tt ACCEPT} (but such ``neighborhood'' is not set by {\tt ACCEPT} per se; a configuration interface will be added when needed). Boundaries of the masked output regions are shown as purple curves in the second and third rows of Figure~\ref{fig:mask_examples}. It is clear from the second row in particular that the combination of the rejection radius {\tt REJECT} and threshold for number of masked neighbors {\tt MASK\_THR} can successfully exclude regions where reconstruction would be unreliable.\footnote{As Confucius said, ``To know what you know and what you do not know, that is true knowledge.'' Translation credit:\\ \url{https://www.brainyquote.com/quotes/confucius_141560}}

In addition to sequestering regions where reliable reconstruction cannot be achieved, {\sc Effortless} also mitigates the impact of masked input pixels in regions that it does reconstruct. Specifically, it ``diffuses'' weights of masked input pixels to their respective four nearest neighbors; each neighbor gets a quarter of the weight. If the masked pixel is on the boundaries of the input pixel window, the weight(s) going to its unselected neighbor(s) is (are) simply abandoned; this is reasonable as outer regions of the weight fields have much smaller absolute values (again, see the third row of Figure~\ref{fig:limit_all}). To ensure the efficiency of the mitigation, such diffusion is agnostic to whether those four nearest neighbors are available and simply done repeatedly. In other words, if a masked pixel gets some weight from its neighbor, it diffuses that additional weight along with its original weight to its own neighbors, including the one giving it the additional weight in the first place. Most of the weights originally carried by lost pixels go to available pixels in the long run, even for clumps of lost pixels.

\begin{figure*}
    \centering
    \includegraphics[width=\textwidth]{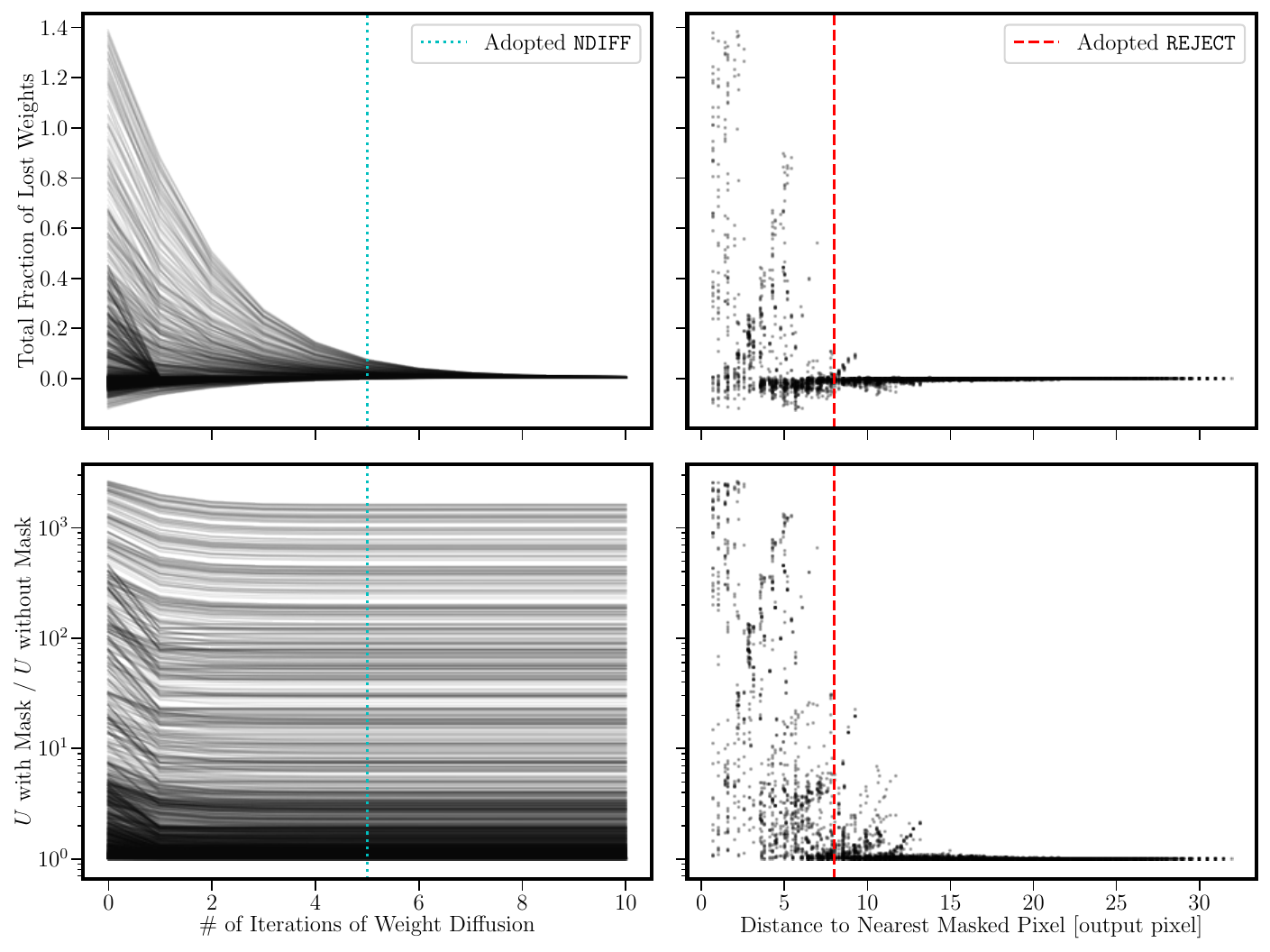}
    \caption{Iterative diffusion of lost weights. The two rows present the total fractions of lost weights and the ratios between PSF leakages with and without masks, respectively. Note that the ``fraction'' can exceed $1$ because some of the weights are negative, hence it is possible for a partial sum to be larger than the total sum. The left column plots these two quantities versus the number of iterations of weight diffusion. The right column plots the initial values of these two quantities versus the distance between each output pixel and its nearest masked input pixel. The adopted number of iterations and rejection radius are shown in the left and right columns as cyan dotted and red dashed vertical lines, respectively.
    \label{fig:mask_diff}}
\end{figure*}

The results achieved by the iterative diffusion of lost weights are illustrated in Figure~\ref{fig:mask_diff}. Absolute values of total fractions of lost weights drop exponentially with the number of iterations, as expected; the PSF leakages decrease to different extents after the first iterations, and gradually converge to final values after a few more iterations. Although the iterative diffusion does help, the final PSF leakages are largely determined by the initial values, which are plotted against distances to the nearest masked input pixels in the lower right panel of Figure~\ref{fig:mask_diff}. As indicated by the red dashed vertical line, the rejection radius of $8$ output pixels can substantially reduce the upper limit of PSF leakages. The hyperparameter {\tt NDIFF} sets the number of iterations and is chosen to be $5$ in this work (shown as cyan dotted vertical lines in the left column). {\sc Effortless} also provides the boolean hyperparameter {\tt RENORM}, which sets whether to renormalize the total weights to values when there are no masked input pixels; but since the total fractions of lost weights rapidly converge to $0$, this option does not matter and is turned off for simulations in this work.

\begin{figure*}
    \centering
    \includegraphics[width=0.9\textwidth]{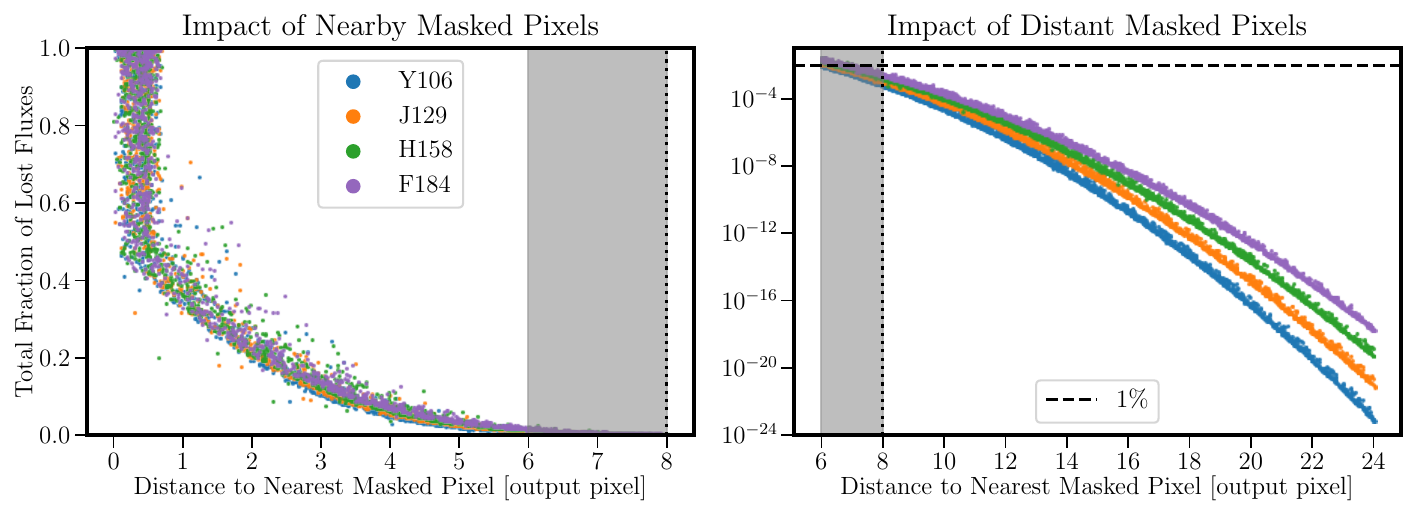}
    \caption{Impact of masked output pixels on injected stars. Both panels plot the total fractions of lost fluxes (assuming the Gaussian target output PSFs are perfectly reconstructed) versus the distances to the nearest masked output pixels. Results for the Y106, J129, H158, and F184 bands are shown in blue, orange, green, and purple, respectively. The $y$-axes of the left and right panels are in linear and logarithmic scales to better display inner and outer features, respectively; the overlapping domain $[6, 8]$ is shown as a gray band in each panel, with its right boundary (the distance threshold adopted in this work) enhanced by a black dotted vertical line. In the right panel, the fraction of $1\%$ is shown as a black dashed horizontal line to aid visual inspection.
    \label{fig:mask_dist}}
\end{figure*}

After the reconstruction of signals in unmasked output pixels, the next issue to consider is the selection of objects. The {\tt \textquotesingle gsstar14\textquotesingle} layer contains $957$ ideal point sources injected on all {\sc HEALPix} nodes with ${\tt NSIDE} = 14$ that are located within the simulated regions \citep[see Figure~1 of ][]{2026ApJ...998..304C}. These injected stars correspond to a representative sample of positions in constructed images, and how they are affected by output pixel masks is shown in Figure~\ref{fig:mask_dist}. Specifically, assuming the Gaussian target output PSFs are perfectly reconstructed, it plots the total fraction of lost fluxes of the injected stars versus their distances to nearest masked output pixels. When the central pixels are masked, it is basically impossible to extract any information about the sources; as the minimum distance increases, the lost fraction decreases, and when the minimum distance is sufficiently large, the lost fraction roughly traces 1D profiles of the corresponding Gaussian functions (note that different bands have different target output PSF widths).

\begin{figure*}
    \centering
    \includegraphics[width=0.9\textwidth]{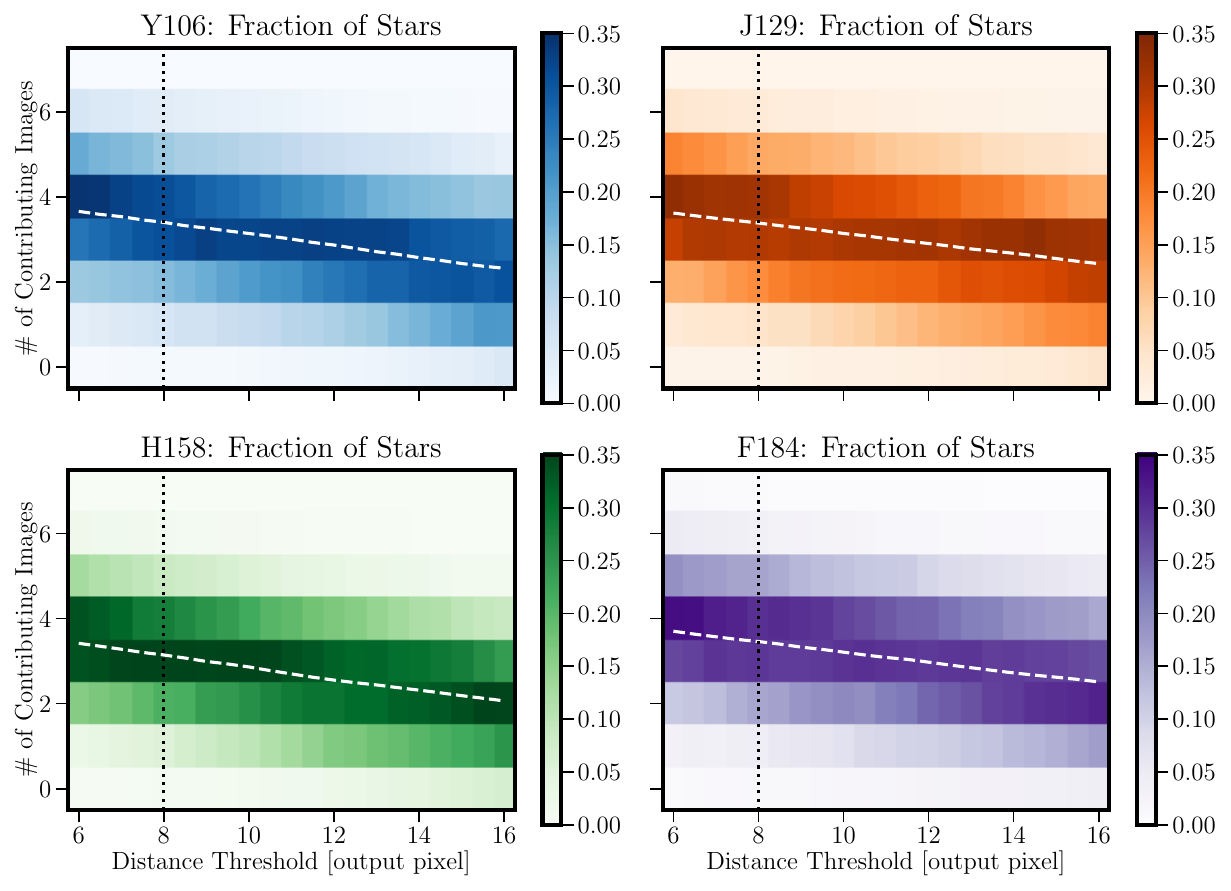}
    \caption{Impact of the threshold for the distances between injected stars and their nearest masked output pixels. Each panel visualizes the distributions of numbers of contributing images as a function of the distance threshold in one of the bands. The average numbers of contributing images at different thresholds are shown as white dashed curves. The distance threshold adopted in this work is shown as a black dotted vertical line in each panel.
    \label{fig:mask_count}}
\end{figure*}

Given this (mostly) monotonic relationship, it is reasonable to set a threshold for the distance to nearest masked output pixel.  With a higher threshold, one can be more confident about the measured properties; but meanwhile, raising the threshold reduces the number of measurable objects. The latter relationship is illustrated in Figure~\ref{fig:mask_count}. In each of the four bands, when the threshold is $\sim 6$ output pixels, the bulk of injected stars can be measured in $\sim 4$ reconstructed images; when the threshold is $\sim 10$, that number drops to $\sim 3$; when it reaches $\sim 16$, the typical number drops to $\sim 2$, and the fraction of stars that cannot be measured in any of the reconstructed images gradually becomes significant. Balancing relationships shown in Figures~\ref{fig:mask_dist} and \ref{fig:mask_count}, the hyperparameter {\tt DISTTHR}\footnote{The usage of underscores in hyperparameter names is unfortunately not completely consistent.} is chosen to be $8$ output pixels for analysis in this work,\footnote{In \citet{2026arXiv260706674C}, I chose ${\tt DISTTHR} = 7.6$ output pixels so that ``measurements of each injected star are provided by at least $1$ image.'' As the area of analysis becomes larger ($1$ block of $(1.75 \,{\rm arcmin})^2$ in one band in that paper, $16$ such blocks in four bands in this paper), having objects that cannot be measured in any of the reconstructed images is inevitable, and we will have to deal with such depth variation.} which coincides with the rejection radius {\tt REJECT}. Such a threshold is shown as black dotted vertical lines in all panels of both figures.

\section{Post-Measurement Calibration} \label{sec:calibr}

After image reconstruction and source selection, the next step is naturally measurement. Following the {\sc Imcom} methodology \citep[see Section~2 of][for a concise summary]{2026ApJ...998..304C}, here I conduct a series of moment-based measurements on the injected stars using the {\sc HSM} \citep{2003MNRAS.343..459H, 2005MNRAS.361.1287M} module of {\sc GalSim} \citep{2015A&C....10..121R}. The measured properties include amplitude $A$ (zeroth moment), centroid offset ${\boldsymbol d}$ (first moments), shear invariant width $s$ and ellipticity $g_1$ and $g_2$ (based on second moments), and the spin-$2$ fourth moment $M^{\rm (4)}_{\rm PSF}$ \citep{2023MNRAS.520.2328Z, 2023MNRAS.525.2441Z}.

\begin{figure*}
    \centering
    \includegraphics[width=\textwidth]{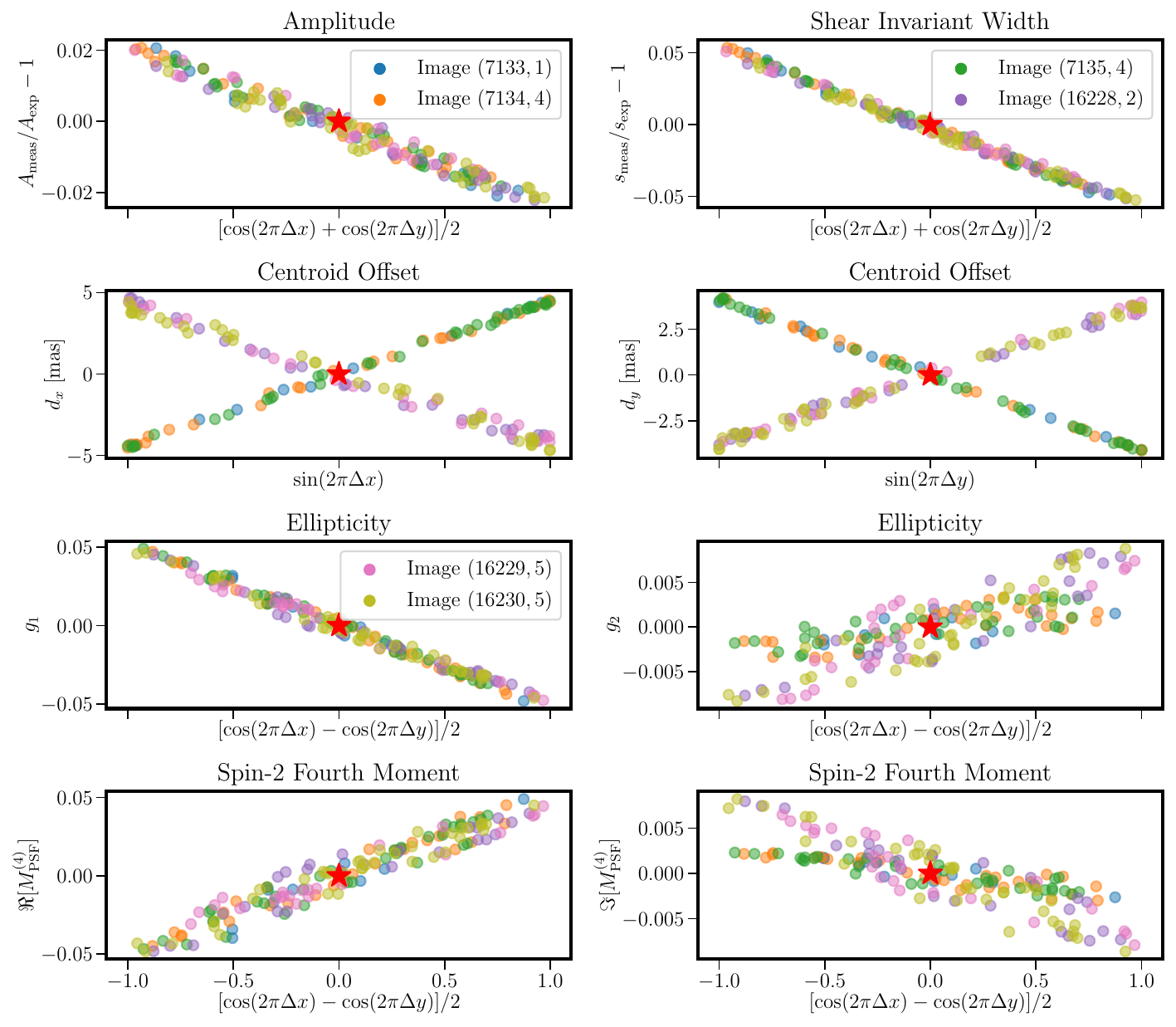}
    \caption{Motivation for post-measurement calibration. Each panel plots a property of injected stars versus some trigonometric function (as shown in the $x$-axis label) of their subpixel positions in the respective input images. From top to bottom, the properties are: amplitude and shear invariant width, centroid offset ($x$- and $y$-components), ellipticity ($2$ components), and the spin-$2$ fourth moment (real and imaginary parts). In each panel, uncalibrated measurements are shown as colored data points; results shown in this figure are taken from block $(14, 12)$ in the Y106 band, same as Figure~3 in my companion paper \citep{2026arXiv260706674C}. In each panel, the $(0, 0)$ point is denoted as a red star, which the distributions usually pass through. The actual calibration makes use of multiple trigonometric functions for each measured property.
    \label{fig:cal_motive}}
\end{figure*}

For all these quantities, the raw {\sc HSM} measurements have relatively large errors. As shown in Figure~\ref{fig:cal_motive}, for this particular block (output image) in the Y106 band (where undersampling of native images is the most severe), astrometric errors are at the ${\rm mas}$ level, fractional errors of the amplitude and the width are at the percent level; errors of $g_1$ and $\Re [M^{\rm (4)}_{\rm PSF}]$ are at the $10^{-2}$ level, while errors of $g_2$ and $\Im [M^{\rm (4)}_{\rm PSF}]$ are at the $10^{-3}$ level. Such large errors certainly exceed Roman weak lensing requirements. Nevertheless, from Figure~\ref{fig:cal_motive}, one can easily see that each of these errors is strongly correlated with some trigonometric function of the subpixel positions. While the slopes differ because of different roll angles, the relationships for each input image are quite robust. Such robust relationships rely on the simple patterns followed by PSF residuals due to finite sampling (see Section~\ref{ss:samp}).

This motivates a post-measurement calibration. Specifically, for each combination of input image and measured property, {\sc Effortless} uses {\sc scikit-learn} {\tt TheilSenRegressor}\footnote{\url{https://scikit-learn.org/stable/modules/generated/sklearn.linear_model.TheilSenRegressor.html}} to fit a relationship between measurement errors and a set of trigonometric functions of subpixel positions of injected stars. The trigonometric functions include $\operatorname{trig} (k \cdot 2\pi \Delta x)$ and $\operatorname{trig} (k \cdot 2\pi \Delta y)$ for $\operatorname{trig} \in \{ \cos, \sin \}$ and $k \in {\mathbb Z} \cap [1, k_{\max}]$ ($4 k_{\max}$ functions in total). To obtain the ``ground truth,'' each reconstructed image has a no-mask counterpart, produced by turning the boolean hyperparameter {\tt NOMASK} on; note that even for real observations, such counterparts can always be produced, as we only need the layers of injected sources. The integer $k_{\max}$ corresponds to the {\sc Effortless} hyperparameter {\tt KMAX} and is set to $2$ for this work after some testing. To prevent overfitting and data leakage, input images with less than $20$ measurable injected stars in the no-mask version are excluded from analysis in \citet{2026arXiv260706674C} and this work. This slightly shifts the distributions in Figure~\ref{fig:mask_count} downward (i.e., reduces the number of usable input images for the injected stars), but note that such exclusion is not caused by a fundamental difficulty, as in principle one can increase the density of injected stars (not necessarily in the same layer to prevent blending). Since the correction depends on properties being corrected (in this case astrometry), an iterative scheme like that in \citet{2025ApJ...990L..26C} is needed; through testing, I find that $\sim 4$ iterations suffice.

\begin{figure*}
    \centering
    \includegraphics[width=\textwidth]{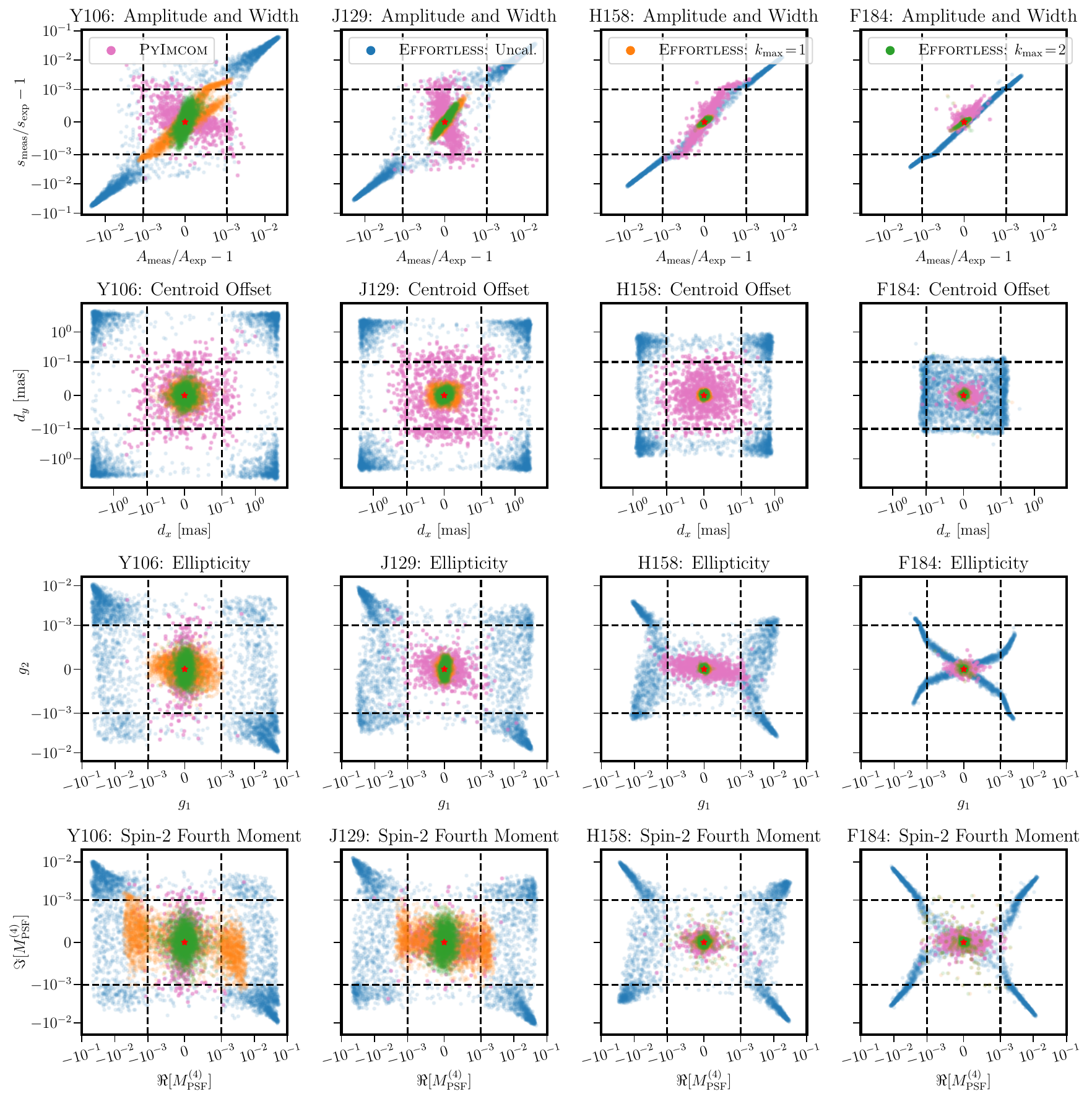}
    \caption{Effect of post-measurement calibration. This figure is an extended version of Figure~3 in my companion paper \citep{2026arXiv260706674C}. From left to right, the four columns correspond to the Y106, J129, H158, and F184 bands, respectively. The mapping between rows and (pairs of) measured properties is the same as in Figure~\ref{fig:cal_motive}. In each panel, {\sc PyImcom} results (based on combined images) are shown in pink, and the uncalibrated, $k_{\max} = 1$ calibrated, and $k_{\max} = 2$ calibrated {\sc Effortless} results (based on individual images) are shown in blue, orange, and green, respectively. Like in Figure~3 of \citet{2026arXiv260706674C}, symmetric logarithmic scales are used; the linear thresholds are different, but they are still shown as vertical and horizontal black dashed lines.
    \label{fig:cal_effect}}
\end{figure*}

After the post-measurement calibration with $k_{\max} = 2$, the distributions of data points in Figure~\ref{fig:cal_motive} are shown in Figure~3 of \citet{2026arXiv260706674C}. For most measured properties of injected stars, the errors are reduced by about $2$ orders of magnitude; see Section~3 of that paper for specific quotes. More comprehensively, the effect of post-measurement calibration is shown in Figure~\ref{fig:cal_effect}. Like in Figure~3 of \citet{2026arXiv260706674C}, the $8$ quantities are grouped into $4$ pairs; with the exception of astrometry (centroid offset), uncalibrated measurements (blue data points) display strong correlations between each pair, with two different coefficients for ellipticity and the spin-$2$ fourth moment. The bimodal distributions are likely associated with the two roll angles in each band of OpenUniverse2024 images used in this work; the two components of the centroid offset do not appear correlated to each other as they are correlated with different trigonometric functions (see the second row of Figure~\ref{fig:cal_motive}; note that the mapping between $\{ \Delta x, \Delta y \}$ and $\{ d_x, d_y \}$ can be different for different images).

Measurements based on combined images produced by {\sc PyImcom} \citep[pink data points; taken from][]{2026ApJ...998..304C} have significantly smaller errors than uncalibrated measurements based on individual images reconstructed by {\sc Effortless}. The correlations between fractional errors of amplitude $A$ and width $s$ and square distributions of centroid offset components indicate that in principle, {\sc PyImcom} results can also be calibrated using subpixel positions; however, since the input images have been combined, how to perform such calibration is not clear. With a simple $k_{\max} = 1$ calibration, {\sc Effortless} errors (orange data points; sometimes hidden behind green ones) are already smaller than or comparable to {\sc PyImcom} errors in most cases. The only exceptions are $g_1$ in the Y106 band and $\Re [M^{\rm (4)}_{\rm PSF}]$ in the Y106 and J129 bands; meanwhile, the $k_{\max} = 1$ calibrated amplitude $A$ and width $s$ still have strong correlations in those two bands. These indicate that at least in the bluer bands, including the $k = 2$ trigonometric functions is beneficial; such speculation is verified by the smaller errors of $k_{\max} = 2$ calibrated {\sc Effortless} results (green data points).

In general, the post-measurement calibration scheme seems successful. Nonetheless, it has several limitations. First, from a practical perspective, it requires producing no-mask counterparts of all images, which more than doubles (since the no-mask version has more unmasked output pixels) the computational costs. Second, from a theoretical perspective, such calibration is purely empirical and lacks a solid theoretical basis (although the patterns followed by PSF residuals caused by finite sampling provide some hints; see Section~\ref{ss:samp}). Third, to generalize such empirical calibration scheme from point sources to extended sources, a much larger ``ground truth'' library needs to be built, since it takes more parameters to describe extended sources (e.g., sizes and ellipticities in addition to positions). Fourth, when the signal-to-noise ratio is low, how the efficacy deteriorates is not clear. Addressing these limitations is an important direction for future work, as I will further discuss in Section~\ref{sec:disc}.

\begin{figure*}
    \centering
    \includegraphics[width=\textwidth]{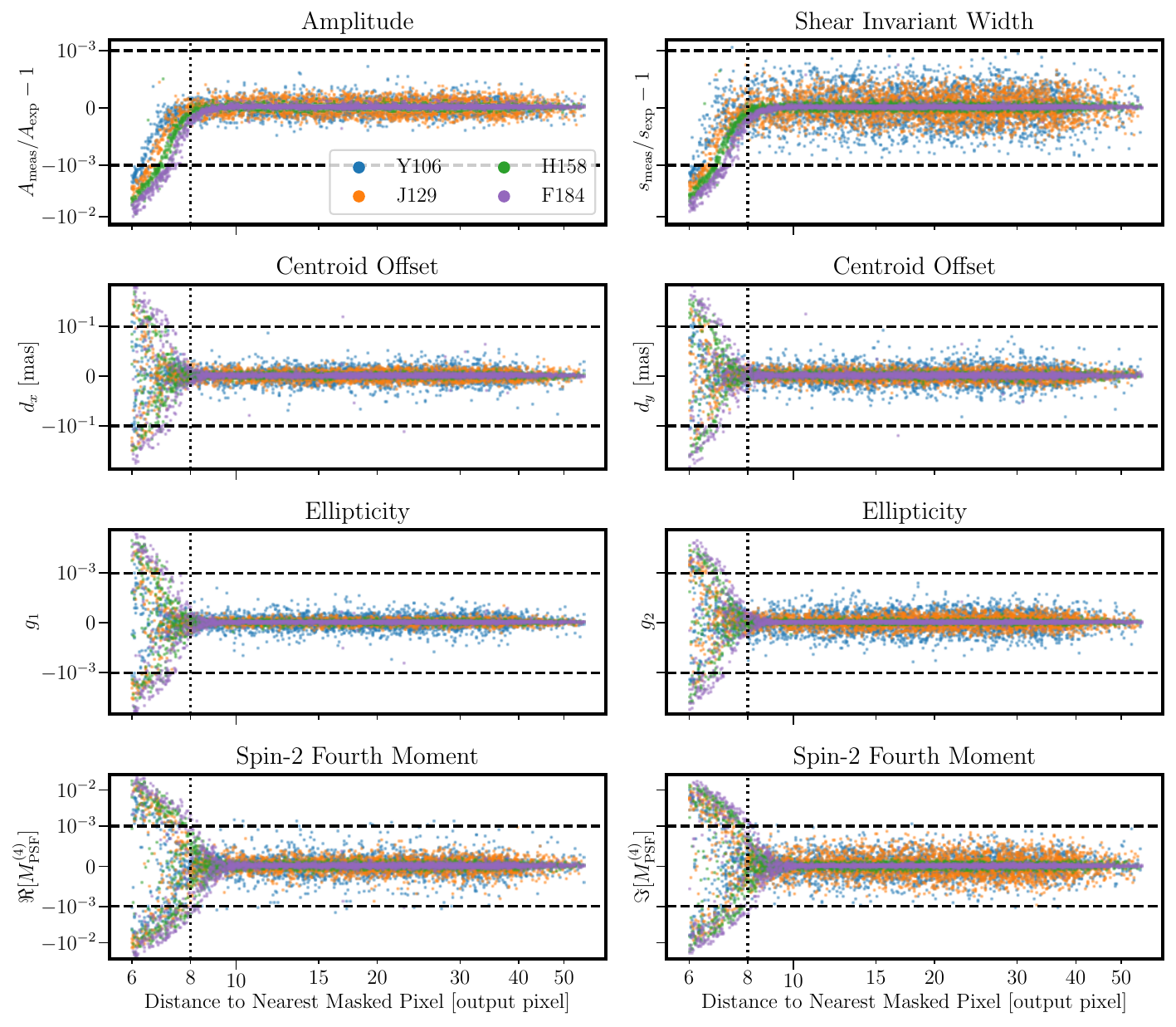}
    \caption{Impact of masked output pixels on measurements of injected stars. The mapping between panels and measured properties is the same as in Figure~\ref{fig:cal_motive}. Each panel plots measured values against the distances between injected stars and their nearest masked output pixels. For the measured properties, the symmetric logarithmic scales are the same as in Figure~\ref{fig:cal_effect}, with the same linear thresholds shown as horizontal black dashed lines. Results for the Y106, J129, H158, and F184 bands are shown in blue, orange, green, and purple, respectively. The distance threshold adopted in this work is shown as a black dotted vertical line.
    \label{fig:cal_mask}}
\end{figure*}

Now that measurements have been performed and calibrated, we can review the threshold for the distance between an injected star and its nearest masked output pixel (i.e., the hyperparameter {\tt DISTTHR}). Figure~\ref{fig:cal_mask} plots $k_{\max} = 2$ calibrated {\sc Effortless} results against such distances. These scatter plots largely buttress what is shown in Figure~\ref{fig:mask_diff}: Objects with nearby masked output pixels are likely to have large errors, but beyond some threshold (like the $8$ output pixels chosen in this work) the masked pixels no longer matter. Interestingly, at small distances to masked output pixels, most measured properties of injected stars manifest a bifurcation feature, but amplitude $A$ and shear invariant width $s$ are monotonic (see the first row of Figure~\ref{fig:cal_mask}). This is because masked output pixels are filled with zero, hence the corresponding parts of the total fluxes are not captured in the image cutouts; as for the width, the size of a bivariate Gaussian function Equation~(\ref{eq:gauss}) is probably underestimated if it is ``squeezed'' by some pixel mask. While the monotonic errors may be calibrated, the bifurcating errors probably cannot. However, it is feasible to fill the masked pixels with signals from other images. Although this would corrupt the simple patterns of PSF residuals, it is possible that intermediate (i.e., between inner and outer) regions are less important in that regard. Exploring such possibility is left for future work.

\begin{figure*}
    \centering
    \includegraphics[width=\textwidth]{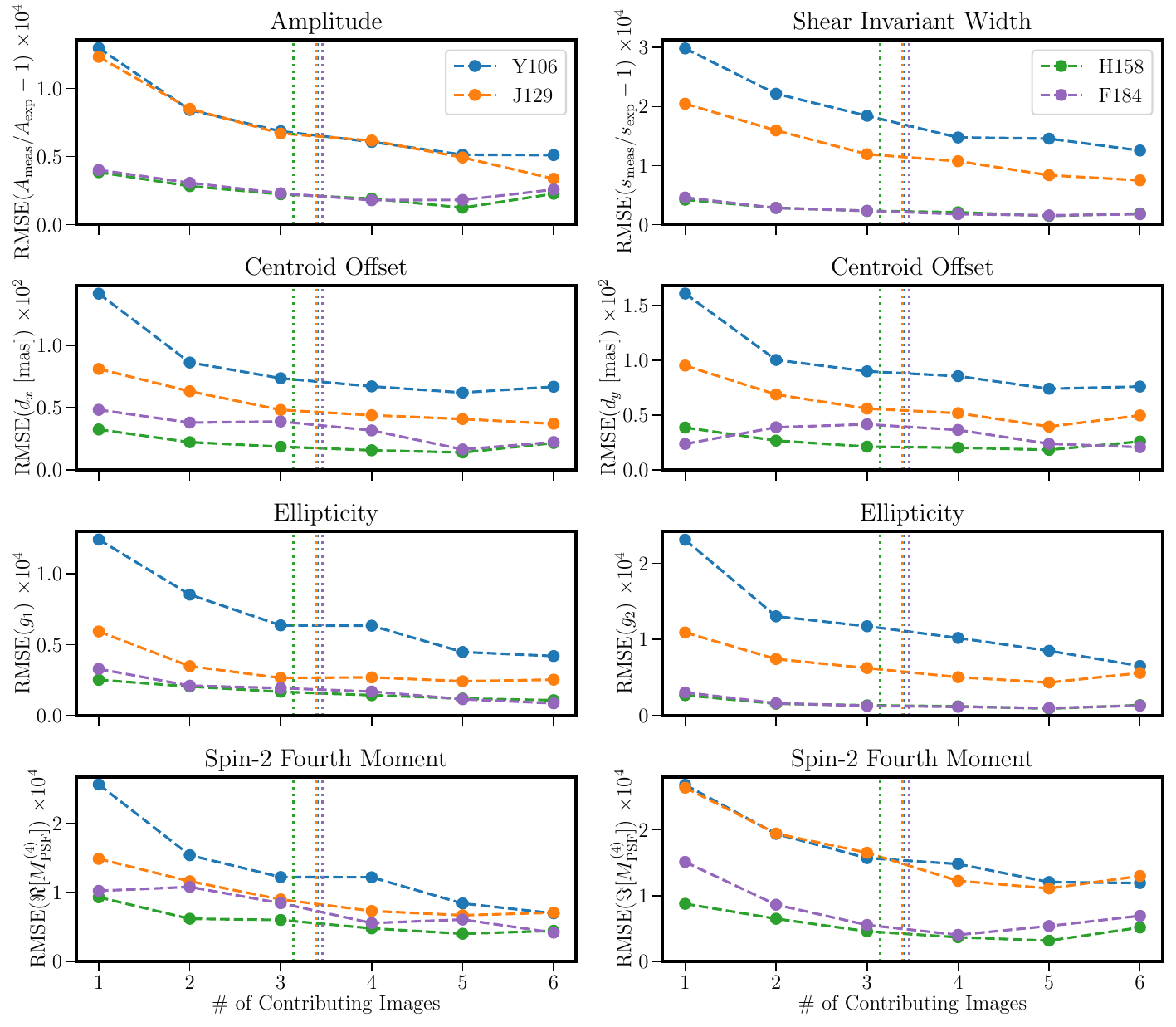}
    \caption{Impact of the number of contributing images. The mapping between panels and measured properties is the same as in Figures~\ref{fig:cal_motive} and \ref{fig:cal_mask}. Each panel plots root mean square errors (RMSEs) as a function of the number of contributing images. Results for the Y106, J129, H158, and F184 bands are shown in blue, orange, green, and purple, respectively. Average numbers of contributing images in different bands, given the distance threshold adopted in this work (see Figure~\ref{fig:mask_count}), are shown as dotted vertical lines.
    \label{fig:cal_count}}
\end{figure*}

Another topic related to the complementarity between different images is the combination of measurement results. While combining images makes PSF residuals hard to track and errors hard to correct, taking the average of available measurements for each object is a reasonable idea. Since errors from different images are not correlated with each other (at least for point sources), the averages are expected to have smaller errors. Such expectation is met in Figure~\ref{fig:cal_count}, with a few exceptions due to small-number statistics when the number of contributing images is very small (like $1$) or very large (like $6$). In most cases, averaging $2$ measurements reduces the errors by a factor of up to $\sim 1.5$, and the gain of increasing the number to $3$ is still significant; with $4$ or more contributing images, the marginal gain becomes smaller. Since {\sc Effortless} seems to allow us to conduct accurate (i.e., unbiased) measurements on individual native Roman images, it is worth considering if we can change the survey strategy (currently $3$ dithers $\times 2$ passes), reduce the number of exposures for each pointing, and cover a larger area of the sky to get better statistics for cosmology and enhance synergies with other probes \citep[e.g.,][]{2024arXiv241104088E}. Results shown here indicate that even if the survey strategy can be changed, we should probably ensure that most objects can be measured in at least $2$ different images. Further investigations along this line are left for future work.

\section{Summary and Discussion} \label{sec:disc}

{\sc Effortless} \citep[EFFicient Optimal image ReconsTruction using LESS memory; previously known as Fast {\sc Imcom};][]{2026AJ....171..140C, 2026arXiv260706674C}, successor to {\sc Imcom} \citep{2011ApJ...741...46R, 2024MNRAS.528.2533H, 2024MNRAS.528.6680Y, 2025ApJS..277...55C, 2026ApJ...998..304C}, is a new algorithm designed for image reconstruction with control over point spread functions (PSFs) in output images. Not only is it tens of times faster than {\sc Imcom}, it can also lead to more accurate measurements for weak gravitational lensing cosmology \citep{2026arXiv260706674C}. Thanks to its efficiency and optimality, {\sc Effortless} has the potential to benefit studies of static features and dynamic aspects of the Universe alike.

In this paper, I have presented the implementation of {\sc Effortless} in detail. I have reviewed the mathematical formalism from a practical perspective (Section~\ref{ss:math_gen}) and discussed the treatment of geometric distortions (Section~\ref{ss:distort}). Importantly, with oversampled native PSFs known a priori, our ability to reconstruct undersampled images is not limited by the Nyquist--Shannon sampling theorem. {\sc Effortless} has a more convenient object-oriented framework, a more general and flexible data model, and a more user-friendly I/O interface (Section~\ref{ss:struct}). Furthermore, it has easily customizable hyperparameters in Python scripts and Jupyter notebooks (Section~\ref{ss:params}) to support varied use cases, including those beyond weak lensing cosmology and beyond the Roman Space Telescope.

I have also addressed many practical issues involved in image reconstruction with {\sc Effortless}. Numerical artifacts when dividing PSFs in Fourier space can be largely eliminated by applying circular bandlimits (Section~\ref{ss:limit}). Finite sampling effects follow relatively simple patterns (Section~\ref{ss:samp}), and can be substantially reduced via a simple post-measurement calibration based on subpixel positions of objects (Section~\ref{sec:calibr}). Finiteness of input pixel windows needs to be balanced with computational costs of image reconstruction (Section~\ref{ss:accept}). {\sc Effortless} uses simple but effective strategies to sequester and mitigate the impact of some input pixels being unavailable (Section~\ref{sec:mask}). Averaging over available measurements from different images can further reduce the errors (Section~\ref{sec:calibr}).

It is worth emphasizing that the precise and accurate measurements presented in \citet{2026arXiv260706674C} and this paper rely on the post-measurement calibration mentioned above, but the generalization of such calibration from point sources (like injected stars) to extended sources (like observed galaxies) may encounter some fundamental difficulty. Overcoming the limitations of the Nyquist--Shannon sampling theorem may necessitate knowing a priori the morphology of the source, which implies additional relations between the Fourier modes that alias to each other that are not available for generic functions (like galaxy profiles). My hope is that, just like we do not know exact positions of injected stars from initial (uncalibrated) measurements, shapes of observed galaxies can be iteratively corrected using a ``truth library'' mentioned in Section~\ref{sec:calibr}. If this does not work, there are two potential fallback strategies. First, instead of calibrating measurements on individual galaxies, we can try to conduct a statistical calibration on each tomographic bin, so that cosmological results are still unbiased. Second, we can combine individual images reconstructed by {\sc Effortless}; leveraging our knowledge about PSF residuals, it should be possible to largely cancel them out. See \citet{2026AJ....171..140C} for discussions about how to combine {\sc Effortless}-reconstructed images.

In any case, this paper is a foundational work bridging the past and the future of {\sc Effortless} development. After mathematical proof-of-concept \citep{2026AJ....171..140C} and promising first results \citep{2026arXiv260706674C}, a practical implementation of {\sc Effortless} is a milestone marking its journey toward maturity. Meanwhile, as mentioned in Section~\ref{sec:calibr}, there is still work to be done before {\sc Effortless} can make significant contributions to astronomical image processing. In addition to exploring potential applications mentioned in the previous two papers, there are fundamental questions to be answered: Can we generalize our understanding of PSF residuals? Can we model noise covariance in reconstructed images? These are crucial topics of follow-up work.

\section*{Acknowledgments}

I appreciate useful discussions with David H. Weinberg and colleagues in the Shear and Clustering Measurement (SCM) Working Group of the Roman High Latitude Imaging Survey (HLIS) Cosmology Project Infrastructure Team (PIT). This paper has undergone internal review in the Roman HLIS Cosmology PIT. I would like to thank Christopher M. Hirata and Mike Jarvis for helpful comments and feedback during the review process.  

This work was supported by the ``Maximizing Cosmological Science with the Roman High Latitude Imaging Survey'' Roman Project Infrastructure Team (NASA grant 22-ROMAN11-0011). I thank Christopher M. Hirata for making the OpenUniverse2024 data needed for this project available on the Ohio Supercomputer Center. The detector mask files used in the Roman image simulations are based on data acquired in the Detector Characterization Laboratory (DCL) at the NASA Goddard Space Flight Center. My colleagues and I would like to thank the personnel at the DCL for making the data available for this project.

Computations for this project used the Cardinal cluster at the Ohio Supercomputer Center \citep{Ohio_Supercomputer_Center2024-dl}. This project made use of the {\sc NumPy} \citep{2020Natur.585..357H}, {\sc Astropy} \citep{2013A&A...558A..33A, 2018AJ....156..123A, 2022ApJ...935..167A}, {\sc Numba} \citep{2015llvm.confE...1L}, {\sc SciPy} \citep{2020NatMe..17..261V}, and {\sc scikit-learn} \citep{2011JMLR...12.2825P} packages. Some of the results in this paper were derived using the {\sc healpy} package \citep{2019JOSS....4.1298Z} based on the {\sc HEALPix} scheme \citep{2005ApJ...622..759G}. Figures in this paper were made using {\sc Matplotlib} \citep{Hunter:2007}; {\sc SAOImageDS9} \citep{2003ASPC..295..489J} played an important role as a preview tool.

\section*{Data Availability}

The OpenUniverse2024 data are publicly available in the NASA/IPAC Infrared Science Archive.\footnote{\url{https://irsa.ipac.caltech.edu/data/theory/openuniverse2024/overview.html}} The {\sc Effortless} (previously known as Fast {\sc Imcom}) and {\sc PyImcom} software packages are available in the following GitHub repositories:
\begin{itemize}
    \item \url{https://github.com/Roman-HLIS-Cosmology-PIT/effortless.git}
    \item \url{https://github.com/Roman-HLIS-Cosmology-PIT/pyimcom.git}
\end{itemize}

Specifically, this work used {\sc Effortless} v0.2.2 and {\sc PyImcom} v1.0.3. The frozen versions are available on Zenodo under an open-source Creative Commons Attribution license: \dataset[doi: 10.5281/zenodo.20365266]{https://doi.org/10.5281/zenodo.20365266} and \dataset[doi: 10.5281/zenodo.17832923]{https://doi.org/10.5281/zenodo.17832923}.

\bibliography{main}{}
\bibliographystyle{aasjournalv7}

\end{document}